\documentclass[10pt,aps,pre,twocolumn,showpacs,notitlepage,floatfix,superscriptaddress,longbibliography]{revtex4-2}

\usepackage{graphicx}
\usepackage{animate}  
\usepackage{dcolumn}
\usepackage{bm}
\usepackage{xcolor}
\usepackage{natbib}
\usepackage{tikz}
\usepackage{pgfplots}
\usepackage{xcolor}
\usepackage{gnuplottex}
\usepackage{amsmath}
\usepackage{float}  
\usepackage{hyperref}
\usepackage[normalem]{ulem}
\usepackage{orcidlink}

\hypersetup{
colorlinks=false,
colorlinks=true,
citecolor=blue,
linkcolor=blue,
urlcolor=blue}

\begin{document}

\preprint{APS/123-QED}
\title{Asymmetry-Induced Chiral Dynamics in Coupled Self-Propelled Robots:  Spinning  and Circular Motion}

\author{Priyanka\orcidlink{0009-0009-8555-2982}}
\email{d24052@students.iitmandi.ac.in}
\affiliation{School of Physical Sciences, Indian Institute of Technology Mandi, Mandi 175001, India}
\author{Nitin Kumar\orcidlink{0000-0003-2697-4842}}
\email{nkumar@iitb.ac.in}
\affiliation{Department of Physics, Indian Institute of Technology Bombay, Powai, Mumbai 400076, India}
\author{Harsh Soni\orcidlink{0000-0003-1189-5543}}
\email{harsh@iitmandi.ac.in}
\affiliation{School of Physical Sciences, Indian Institute of Technology Mandi, Mandi 175001, India}




\date{\today}
\begin{abstract}
Motivated by the chiral motility of microswimmers, we investigate how geometric asymmetry in a system of two self-propelled active Brownian robots coupled by a spring gives rise to rich emergent dynamics. We demonstrate that asymmetry in the robots' propulsion directions generates net torques that induce persistent rotational motion. Depending on the choice of propulsion angles $\alpha_1$ and $\alpha_2$, the system exhibits three distinct dynamical regimes---run-and-tumble motion, circular trajectories, and spinning---with the geometric configuration primarily determining the realized regime. We further show that spring stiffness and rotational noise act as additional tuning parameters governing the stability of these regimes. These results show how the interplay of mechanical coupling and activity produces diverse self-organized dynamics in simple robotic dimers, providing a bridge between artificial active systems and biological microswimmers such as bacteria, \textit{Chlamydomonas reinhardtii}, and spermatozoa.

 \end{abstract}

\maketitle


\section{\label{sec:level1}Introduction}
At microscopic scales, self-propelled microorganisms navigate their environments using a variety of motility strategies that enable them to search for nutrients, avoid unfavorable conditions, and interact with their surroundings \cite{Lauga2009,Elgeti2015}. These strategies include flagellar swimming, in which rotating helical flagella propel bacteria such as \textit{E. coli} \cite{Berg2003}; breaststroke-like flagellar beating in biflagellated algae (e.g., \textit{Chlamydomonas reinhardtii}) \cite{Polin2009}; ciliary propulsion in organisms like \textit{Paramecium} \cite{Brennen1977}; and surface-based motility, including gliding and twitching, in certain bacteria \cite{Jarrell2008}. Although microorganisms employ a wide variety of propulsion strategies, many exhibit stochastic swimming patterns that enhance environmental exploration \cite{Huoetal2021, Berg1993, Cates2012, Fieretal2017}.\\
The most extensively studied example is run-and-tumble (RnT) motion, in which cells alternate between relatively straight ``runs'' and rapid reorientation events, ``tumbles'', that randomize the swimming direction \cite{Berg1972, Tailleur2008, Paramanick2025, Polin2009,Bechinger2016}. In \textit{E. coli}, tumbles arise from transient reversals of flagellar motor rotation causing bundle unbundling \cite{Berg1972, Darnton2007, Berg2003} whereas in \textit{Chlamydomonas reinhardtii}, transitions between synchronized and desynchronized flagellar beating give rise to analogous RnT dynamics \cite{Polin2009}. Beyond stochastic reorientation, these same organisms also display circular trajectories. For instance, \textit{E. coli} swims in clockwise circles near solid boundaries due to hydrodynamic interactions \cite{Lauga2006, DiLuzio2005}. \textit{Chlamydomonas} rotates at 1–2 Hz during swimming, tracing helical paths arising from nonplanar flagellar beats that require symmetry breaking between the two flagella \cite{Ruffer1985, Bayly2010, Cortese2021, Dutcher2019, Wang2026}.

Intriguingly, biological microswimmers actively modulate these internal mechanical parameters to navigate \cite{Friedrich2012}; recent experiments have shown that \textit{Chlamydomonas reinhardtii} switches its circling handedness between counterclockwise and clockwise states by altering its flagellar beating parameters in response to changing light intensities \cite{Wang2026, xin2020optically}. This capacity highlights how complex biological steering is ultimately executed by physical modifications in flagellar asymmetry and net torque \cite{Gadelha2020}. Most microorganisms lack perfect axial symmetry: bacterial flagella attach off-center \cite{Berg2003}, spermatozoa exhibit morphological asymmetry of the head \cite{Gadelha2020}, and \textit{Chlamydomonas} possesses inherent differences between its \textit{cis} and \textit{trans} flagella \cite{Kamiya1987, Wei2024}. Even slight geometric asymmetry couples the propulsion force to a net torque, deflecting otherwise straight trajectories into curved paths \cite{Lauga2009, Kummel2013, Liebchen2022}. Consequently, circular or helical swimming represents the generic mode of self-propulsion, while straight trajectories arise only under special symmetry conditions or are recovered statistically through stochastic reorientation events \cite{Lauga2006, Lauga2009, Polin2009}. 

Motivated by these biological observations, Paramanick et al.~\cite{Paramanick2025} experimentally studied a minimal robotic model consisting of two self-propelled robots connected by a rigid rod, designed to mimic run-and-tumble-like motion of microorganisms. Each robot was actuated by motorized wheels and operated in an overdamped regime \cite{Paramanick2024}, analogous to microscopic swimmers at low Reynolds numbers where inertia is negligible \cite{Purcell1977}. The fixed-length connection between the robots introduced mechanical coupling that converted individual propulsion forces into collective motion. By tuning the pivot offset $\delta$ -- the distance between the rod attachment point and the robot center -- and the propulsion angle $\alpha$ -- the angle between the propulsion direction and the line joining the pivot point to the robot center [Fig.~\ref{fig:schematic}] -- transitions between straight runs and sudden tumbles emerge even in the absence of hydrodynamic interactions \cite{Paramanick2025}. This model demonstrated that simple mechanical constraints can reproduce complex motility patterns characteristic of biological microswimmers \cite{Lauga2009, Elgeti2015, Xia2024}.

While this rigid-rod model successfully replicated RnT motion under symmetric coupling, a key question remains: can controlled mechanical asymmetry generate circular motion and give rise to additional dynamical regimes? Furthermore, real biological systems employ elastic rather than rigid connections, as flagellar basal bodies are linked by compliant fibers \cite{Wan2016, Geyer2013, Bianchi2022, Liu2018}. Yet whether such compliance plays a functional role in switching between motility behaviors  remains less explored.

Here, we address these questions by investigating the dynamics of a spring-coupled pair of self-propelled active Brownian robots with asymmetric propulsion directions, extending the rigid-rod model of Ref.~\cite{Paramanick2025, Zheng2023} to include elastic compliance. We investigate the spinning and circular dynamical regimes that emerge from geometric asymmetry \cite{Kummel2013, Lowen2016, Liebchen2022}. We show that the chirality of the motion---clockwise or counterclockwise---is determined by the sign of $\alpha_1-\alpha_2$, and identify small parameter-space regions where the chirality is reversed due to the presence of alternative stable fixed points. We further show that rotational noise $D_r$ and spring stiffness $K$ modulate these regimes: increasing $D_r$ progressively randomizes the propulsion orientations, weakening the geometric torque and eventually destroying both spinning and circular motion \cite{Caprini2022, Caprini2023,Caprini2026}, while increasing $K$ enhances the dynamical response of both regimes, saturating in the rigid-rod limit. These emergent dynamics originate from geometric asymmetry, without requiring hydrodynamic interactions, chemical gradients, or active sensing \cite{Kummel2013,Liebchen2022,Chan2024}.

The remainder of the paper is organized as follows. Sec.~\ref{model} presents the theoretical model. Sec.~\ref{result} discusses the results. Sec.~\ref{R&D} contains the discussion and conclusions.

\section{THEORETICAL MODEL}~\label{model}
Our theoretical model is motivated by the experimental study of Paramanick et al.~\cite{Paramanick2025}, which investigated the dynamics of a pair of active Brownian robots connected by a rigid rod. Here, we generalize their model by replacing the rigid rod with a flexible Hookean spring, which reduces to a rigid constraint in the limit of a large spring constant. We first present the model for a single active Brownian robot subject to an external force (see Sec.~\ref{singrob}), and then extend it to the case of coupled robots (see Sec.~\ref{cuprob}).
\subsection{Single active Brownian robot}~\label{singrob}

\begin{figure}[t]
\centering
\includegraphics[width=1\linewidth]{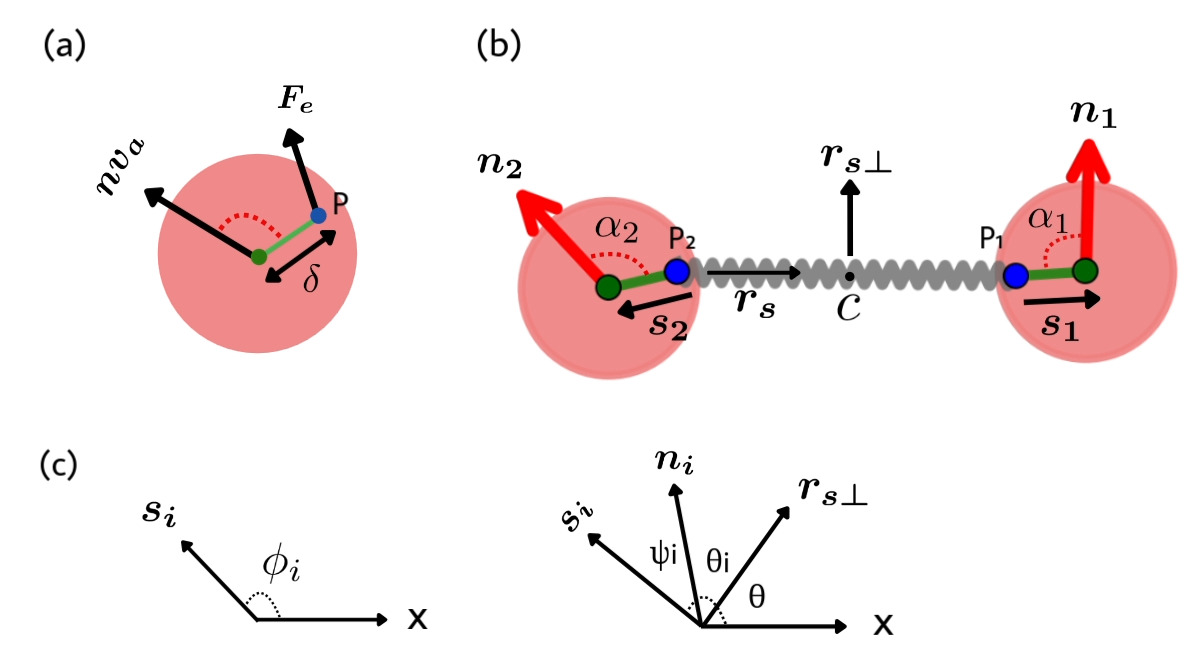}
\caption{Schematic diagram of the system: (a) A single active Brownian robot subjected to an external force $\mathbf{F}_e$ applied at a point $P$, located at a distance 
$\delta$ from its center. (b) Two robots coupled via a Hookean spring. (c) Illustration of a vector showing the various angular coordinates used in the calculation.}
\label{fig:schematic}
\end{figure}

Here, we model a single active Brownian robot subjected to an external force. Let the instantaneous velocity and the $z$ component of the angular velocity be denoted by $\mathbf{v}$ and $\omega$, respectively. The model is based on the following key assumptions: (i) all resistive effects are incorporated into generalized dissipative force and torque, $\mathbf{F}_d = -\boldsymbol{\Gamma} \cdot \mathbf{v}$ and $\tau_d = -\Gamma_\tau \omega$, where $\boldsymbol{\Gamma}$ and $\Gamma_\tau$ are constant matrix and scalar coefficients, respectively;
(ii) the center of mass coincides with the geometric center of the robot;  and (iii) due to its wheeled geometry, the robot experiences anisotropic translational friction, with different resistive responses parallel and perpendicular to its direction of propulsion. This anisotropy is captured by choosing
\begin{equation}
\Gamma = (\Gamma_{\parallel}-\Gamma_{\perp})\,\mathbf{n}\mathbf{n} + \Gamma_{\perp}\,\mathbf{I},
\end{equation}
where $\Gamma_{\parallel}$ and $\Gamma_{\perp}$ denote the translational dissipation coefficients along and perpendicular to the robot’s orientation vector $\mathbf{n}$, respectively, and $\mathbf{I}$ is the identity tensor.
Under these assumptions, the equations of motion for a robot subjected to an external force $\mathbf{F}_e$ applied at a distance $\delta$ from its center are given by [Fig.~\ref{fig:schematic}(a)]


\begin{subequations}\label{singrobeq}
\begin{align}
m\dfrac{d \mathbf{v}}{dt}=&-\mathbf{\Gamma}\cdot\mathbf{v} + F_\mathrm{a}\,\mathbf{n}(t) + \mathbf{F}_e, \\
I \dfrac{d \mathbf{\omega}}{dt}=&- \Gamma_{\tau}\omega  -\,
\mathbf{z}\!\cdot\!(\delta\mathbf{s}\times\mathbf{F}_e)
+ \sqrt{2\Gamma^2_{\tau}D_r}\,\eta(t),
\end{align}
\end{subequations}
where $m$ and $I$ are the mass and the moment of inertia about the center of the robot, respectively. The quantity $F_\mathrm{a}$ denotes the active force generated by the robot’s motor, $\mathbf{s}$ is the direction from the pivot point to the robot center, $\mathbf{z}$ is the unit vector normal to the plane of motion, and $D_r$ is the rotational diffusion coefficient. The stochastic function $\eta(t)$ represents Gaussian white noise with zero mean and unit variance. We further assume that the robot operates in the overdamped limit, allowing us to neglect the inertial terms $md \mathbf{v}/dt$ and $I d \boldsymbol{\omega}/dt$~\cite{Purcell1977}. The equations of motion then reduce to
\begin{subequations}\label{singrobeq}
\begin{align}
\mathbf{v} &= v_\mathrm{a}\,\mathbf{n}(t) + \boldsymbol{\Gamma}^{-1}\!\cdot\!\mathbf{F}_e, \\
\omega &= -\frac{\delta}{\Gamma_{\tau}}\,
\mathbf{z}\!\cdot\!(\mathbf{s}\times\mathbf{F}_e)
+ \sqrt{2D_r}\,\eta(t),
\end{align}
\end{subequations}
where
\begin{equation}
\boldsymbol{\Gamma}^{-1} = \gamma\,\mathbf{n}\mathbf{n} + \frac{1}{\Gamma_{\perp}}\,\mathbf{I}; 
\qquad 
\gamma = \frac{1}{\Gamma_{\parallel}} - \frac{1}{\Gamma_{\perp}},
\end{equation}
and $v_\mathrm{a} = F_\mathrm{a}/\Gamma_{\parallel}$ is the self-propulsion speed of the robot along its orientation in the absence of external forces; we assume it to be constant.
\subsection{Coupled active Brownian robots}~\label{cuprob}
We consider two active Brownian robots connected by a Hookean spring of stiffness $k$.
Denoting the orientation of the $i$th robot by $\mathbf{n}_i$ and the direction from the pivot point P$_i$ to its center by $\mathbf{s}_i$, where $i = 1$ for the right robot and $i = 2$ 
for the left robot, and c is centroid of the system as shown in Fig.~\ref{fig:schematic}(b). The orientation angle $\psi_i$ of $\mathbf{s}_i$ with respect to $\mathbf{n}_i$ is fixed for each robot. The force $\mathbf{F}_i$ on the $i$th robot due to the spring can be expressed in terms of the vector $\mathbf{r}_s$  pointing from P$_2$ to P$_1$.
\begin{equation}
    \mathbf{F}_i = p_i\,k\bigl(r_s - l\bigr)
    \hat{\mathbf{r}}_s,
\end{equation}
where $l$ is the spring's rest length. The sign parameters $p_1 = -1$ and $p_2 = +1$ are chosen to ensure that the forces on the two robots are equal in magnitude and opposite in direction, consistent with Newton's third law. Let $\mathbf{r}_i$ denote the position of the center of the $i$th robot. The pivot point P$_i$ is located at $\mathbf{r}_{si} = \mathbf{r}_i + \delta\,\mathbf{s}_i$, and
\begin{equation}
\mathbf{r}_s = \mathbf{r}_{s1} - \mathbf{r}_{s2}
= (\mathbf{r}_1 + \delta\,\mathbf{s}_1) - (\mathbf{r}_2 + \delta\,\mathbf{s}_2).
\end{equation}
The equations of motion for the position $\mathbf{r}_i$ of the $i$th robot and the orientation angle $\phi_i$ of $\mathbf{s}_i$ follow from Eq.~\eqref{singrobeq} and are given by
 \begin{align}
\frac{d\mathbf{r}_i}{dt} &=
v_a\mathbf{n}_i
 + p_i k(|\mathbf{r_s}| - l)
   \left[
     \gamma\,\mathbf{n}_i(\mathbf{n}_i\!\cdot\!{\mathbf{r_s}})
     + \frac{1}{\Gamma_\perp}{\mathbf{r_s}}
   \right], \\[6pt]
\frac{d\phi_i}{dt} &=
 -p_i\,\frac{\delta k(|\mathbf{r_s}| - l)}{\Gamma_\tau}
   (\mathbf{s}_i\!\cdot\!{\mathbf{r_s}}_\perp)
   + \sqrt{2D_r}\,\eta_i(t)
\end{align}
where $\eta_i(t)$ is Gaussian white noise with zero mean and delta correlation, $\langle \eta_i(t)\eta_j(t') \rangle = \delta_{ij}\,\delta(t - t')$. Furthermore, $\mathbf{z}\!\cdot\!(\mathbf{s}_i\times\mathbf{r}_s) = \mathbf{s}_i\!\cdot\!\mathbf{r}_{s\perp}$, where ${\mathbf{r}}_{s\perp}$ is the unit vector perpendicular to the spring. 

Unless stated otherwise, all simulations are performed using the following parameter values: self-propulsion speed $v_a = 5.0~\mathrm{cm\,s^{-1}}$, rotational diffusion coefficient $D_r = 0.02~\mathrm{rad^2\,s^{-1}}$, dimensionless spring stiffness $K = k/(D_r \Gamma_\parallel) = 3000$, $\Gamma_{\parallel}/\Gamma_{\perp} = 0.1$, equilibrium spring length $l = 14.5~\mathrm{cm}$, pivot offset $\delta = 3.0~\mathrm{cm}$, and robot diameter $d = 7.5~\mathrm{cm}$. These values are chosen to match the experimental conditions of~\cite{Paramanick2025}. Any deviations from these values are explicitly indicated in the figure captions.

\section{Results} \label{result}
The emergence of run-and-tumble dynamics in the symmetric case $\alpha_1 = \alpha_2$ has already been discussed in detail in Ref.~\cite{Paramanick2025}. Here, we focus on how introducing asymmetry into the system, i.e., taking $\alpha_1 \neq \alpha_2$ generates a net torque on the system, giving rise to two additional qualitatively distinct dynamical regimes: \textit{spinning motion}, in which the geometric torque converts the propulsive input into rapid in-place rotation with negligible translation, and \textit{circular motion}, where the interplay between translational and rotational dynamics drives the centroid along sustained curved trajectories with geometry-controlled radius and chirality. 

\begin{figure}[htbp]
    \centering
    \includegraphics[width=0.99\linewidth]{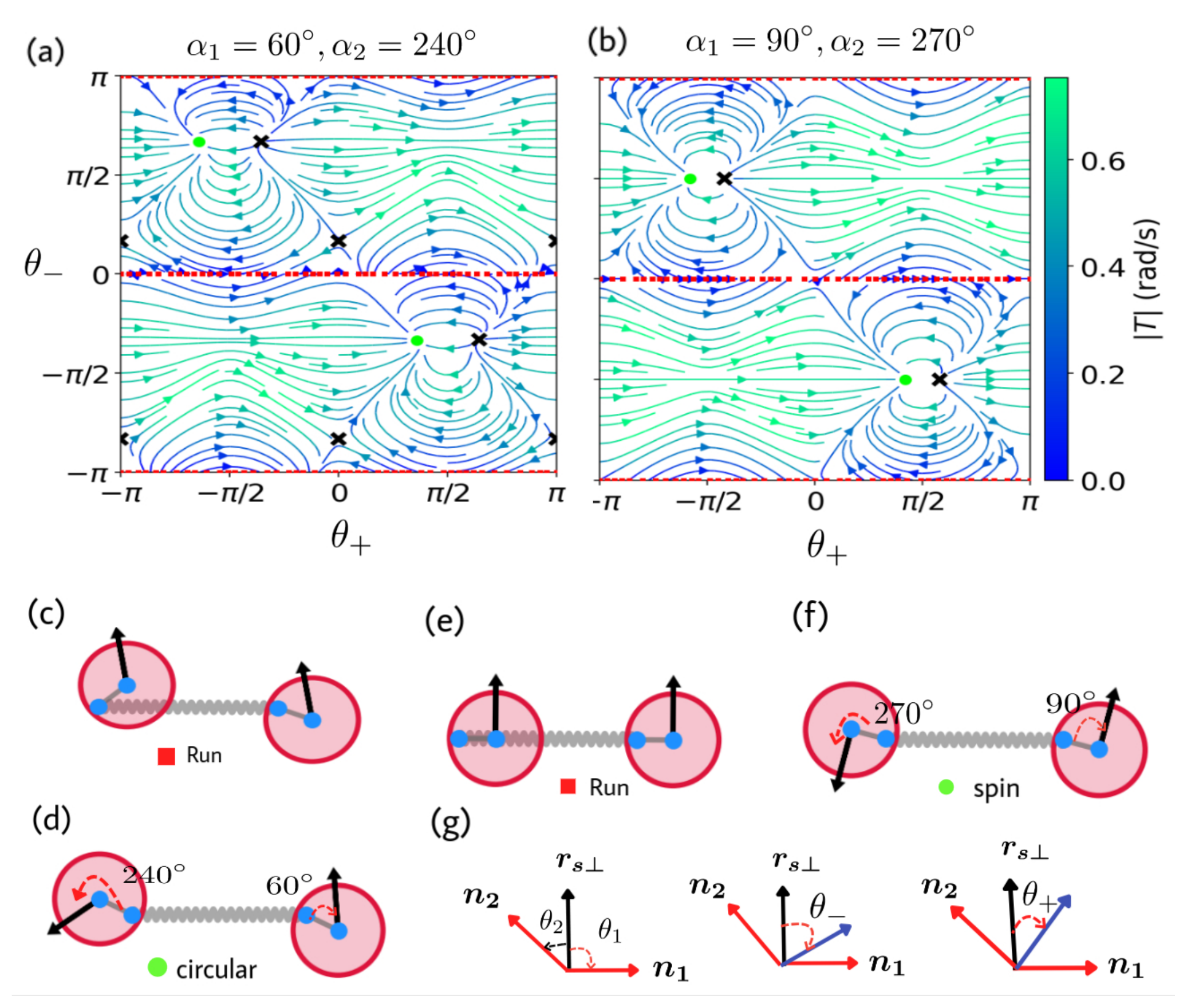}
    \caption{Stream plots for two angular configurations and their corresponding dynamical states.
    The flow in the $(\theta_+, \theta_-)$ plane is shown for (a) $\alpha_1 = 60^\circ$, $\alpha_2 = 240^\circ$  and (b) $\alpha_1 = 90^\circ$, $\alpha_2 = 270^\circ$. 
    The red dashed line at $\theta_- = n\pi$ represents the run state, where $\mathbf{n}_1 \parallel \mathbf{n}_2$, illustrated by the red square configurations below (c),(e). 
    The green dots denote stable fixed points corresponding to circular motion in (d) and spinning motion in (f). (g) Angular coordinates $(\theta_{+}, \theta_{-})$, where $\theta_{\pm}=(\theta_{1}\pm\theta_{2})/2$.}
    \label{fig:stream}
\end{figure}

To understand the origin of the spinning and circular states, we analyze the deterministic flow in the $(\theta_+, \theta_-)$ plane at $D_r = 0$. Here, $\theta_\pm = (\theta_1 \pm \theta_2)/2$, and $\theta_1$ and $\theta_2$ denote the orientations of the two robots relative to the vector $\mathbf{r}_{s\perp}$ as shown in the Fig.~\ref{fig:stream}(g). The deterministic equations of motion for $\theta_\pm$ that govern this flow are~\cite{Paramanick2025}
\begin{equation}
\frac{d\theta_{\pm}}{dt} = T_\pm,
\end{equation}
where
\begin{align}
T_+ &= 2v_a \sin\theta_- 
\left[\frac{\sin\theta_+}{l} - \cos\theta_+ \, \mathcal{G}\, \mathcal{H}\right], 
\label{eq:Tplus} \\[6pt]
T_- &= \frac{2\delta v_a}{\Gamma_\tau}\, \mathcal{G}\, 
\cos(\theta_+ + \psi_+)\cos\theta_+ \cos(\theta_- + \psi_-)\sin\theta_-,
\label{eq:Tminus}
\end{align}
with
\begin{align}
\mathcal{G} &= \frac{1}{\gamma\sum\limits^{2}_{i=1}\sin^2\theta_i
+ \dfrac{\delta^2}{\Gamma_\tau}\sum\limits^{2}_{i=1}\cos^2(\theta_i + \psi_i) 
 + \dfrac{2}{\Gamma_\perp}}, 
\label{eq:G} \\[6pt]
\mathcal{H} &= \frac{\delta}{\Gamma_\tau}\sin(\theta_+ + \psi_+)\sin(\theta_- + \psi_-) 
- \frac{\gamma}{l}\sin(2\theta_+)\cos(2\theta_-) \notag \\
&\quad + \frac{\delta^2}{\Gamma_\tau l}\sin\!\big[2(\theta_+ + \psi_+)\big]
\cos\!\big[2(\theta_- + \psi_-)\big],
\label{eq:H}
\end{align}
and $\psi_\pm = (\psi_1 \pm \psi_2)/2$, with $\psi_1 = \alpha_1 - \pi$ and $\psi_2 = \pi - \alpha_2$.
Fig.~\ref{fig:stream}(a) and (b) illustrate the corresponding flow fields for $(\alpha_1, \alpha_2) = (60^\circ, 240^\circ)$ and $(90^\circ, 270^\circ)$, respectively.
The red dashed line at $\theta_- = n\pi$ represents configurations where $\mathbf{n}_1 \parallel \mathbf{n}_2$, i.e., both robots are aligned in the same direction Fig.~\ref{fig:stream}(c) and (e). The system is stable along these lines with respect to $\theta_-$, giving rise to a run state with the maximum translational speed $v_a$ (see movies~\href{https://drive.google.com/drive/folders/1ZZfnyFyMga9KfV9fWmAVixl5AZbw6R6D?usp=sharing}{S1} and~\href{https://drive.google.com/drive/folders/1ZZfnyFyMga9KfV9fWmAVixl5AZbw6R6D?usp=sharing}{S2}). However, it is unstable with respect to $\theta_+$, indicating that the direction of motion is not fixed during the run state. The stable fixed points (shown as green dots) lead to either circular or spinning motion (see movies~\href{https://drive.google.com/drive/folders/1ZZfnyFyMga9KfV9fWmAVixl5AZbw6R6D?usp=sharing}{S3} and~\href{https://drive.google.com/drive/folders/1ZZfnyFyMga9KfV9fWmAVixl5AZbw6R6D?usp=sharing}{S4}), depending on the angular configuration, as illustrated in Fig.~\ref{fig:stream}(d), (f). For $\alpha_1 + \alpha_2 \neq 360^\circ$, at the stable fixed points, the orientations of the two robots are not perfectly opposite to each other. As a result, the net active force on the system is nonzero. Therefore, the system not only spins about its axis but also drifts, which in turn gives rise to circular motion, as shown in Fig.~\ref{fig:stream}(d). In contrast, when $\alpha_1 + \alpha_2 = 360^\circ$, the stable fixed points correspond to perfectly anti-aligned robots. In this case, the net active force on the system vanishes, and the system undergoes pure spinning motion; the centroid does not execute any circular motion [Fig.~\ref{fig:stream}(f)].

In this work, we focus exclusively on the dynamics near these stable fixed points, as they determine the emergence and stability of the circular and spinning regimes explored throughout the study.

\subsubsection{Spinning motion}
\begin{figure}[t]
\centering

\includegraphics[width=0.49\columnwidth]{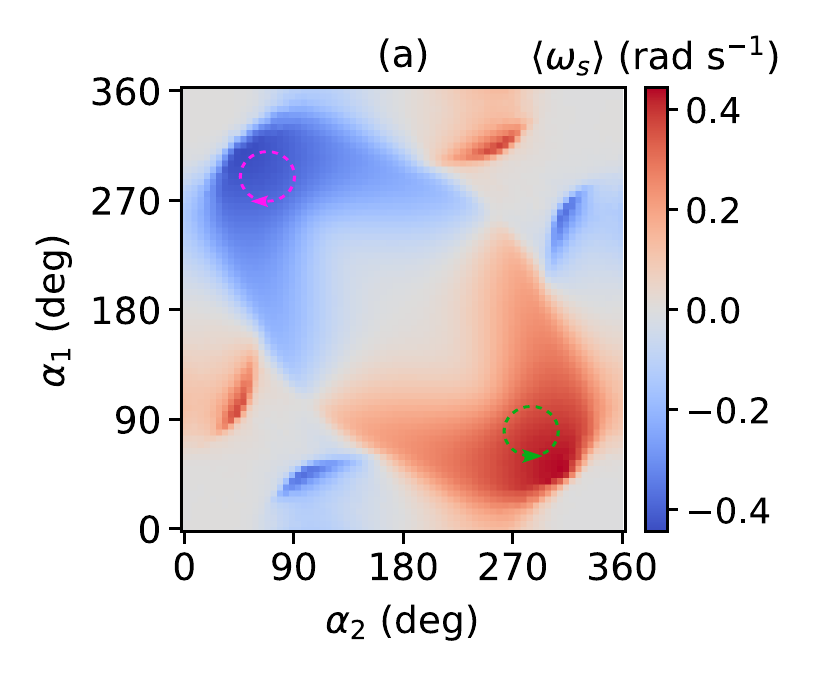}
\includegraphics[width=0.49\columnwidth]{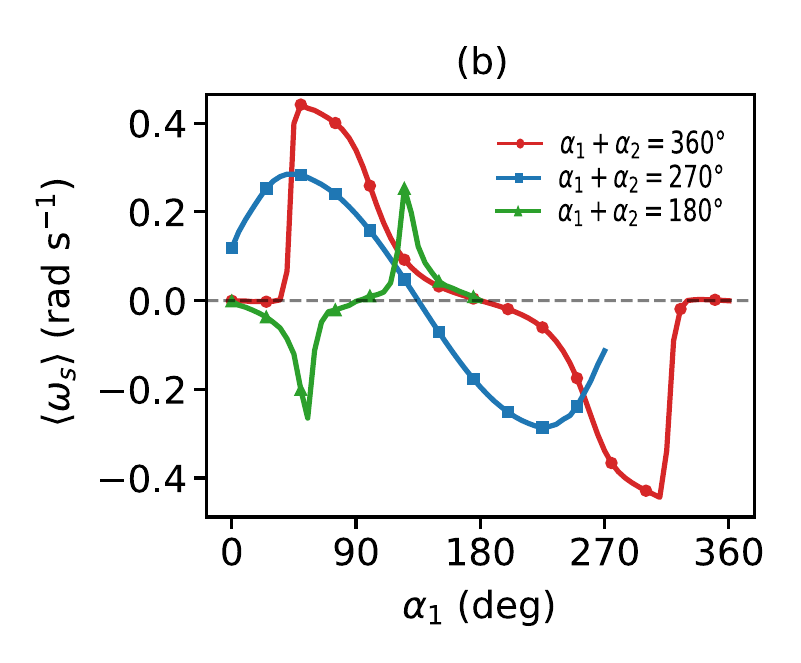}
\caption{(a) Heatmap of the average spin angular velocity $\langle \omega_s \rangle$ in the $(\alpha_1,\alpha_2)$ parameter space. Positive and negative values correspond to counterclockwise and clockwise rotation, respectively. (b) Corresponding one-dimensional profiles of $\langle \omega_s \rangle$ along the anti-diagonal lines $\alpha_1+\alpha_2=360^\circ$, $270^\circ$, and $180^\circ$.}
\label{fig:combined_plots}
\end{figure}

To quantify the spinning motion of the coupled robot, we define its spin angular velocity $\omega_s$ as the angular velocity of the line joining the two robots i.e., the connecting spring.
Fig.~\ref{fig:combined_plots}(a) shows the heat map of the average spin angular velocity $\langle \omega_s \rangle$ as a function of the angles $\alpha_1$ and $\alpha_2$.
For $\alpha_1 = \alpha_2$, the propulsion geometry is symmetric, and the system exhibits no systematic rotation, as reflected by $\langle \omega_s \rangle = 0$ in the heatmap. The dynamics are therefore predominantly translational. 

Away from this symmetric line, geometric asymmetry induces a finite active torque that results in persistent spinning motion. The blue and red regions of the heat map indicate clockwise and counterclockwise motion, corresponding to positive and negative values of $\langle \omega_s \rangle$, respectively (see movies~\href{https://drive.google.com/drive/folders/1ZZfnyFyMga9KfV9fWmAVixl5AZbw6R6D?usp=sharing}{S5} and~\href{https://drive.google.com/drive/folders/1ZZfnyFyMga9KfV9fWmAVixl5AZbw6R6D?usp=sharing}{S6}). The effect is most pronounced along the line $\alpha_1+\alpha_2=360^\circ$, where the propulsion directions of the two robots are exactly opposite. However, not all configurations on this line produce spinning. In particular, for $\alpha_1=\alpha_2=180^\circ$, the propulsion directions are anti-aligned along the spring axis, resulting in zero net torque and consequently no spinning motion (see movie~\href{https://drive.google.com/drive/folders/1ZZfnyFyMga9KfV9fWmAVixl5AZbw6R6D?usp=sharing}{S7}).

\begin{figure}[t]
\centering
\begin{minipage}{0.49\columnwidth}
    \centering
    {\footnotesize{(a)}}
    \includegraphics[width=\linewidth]{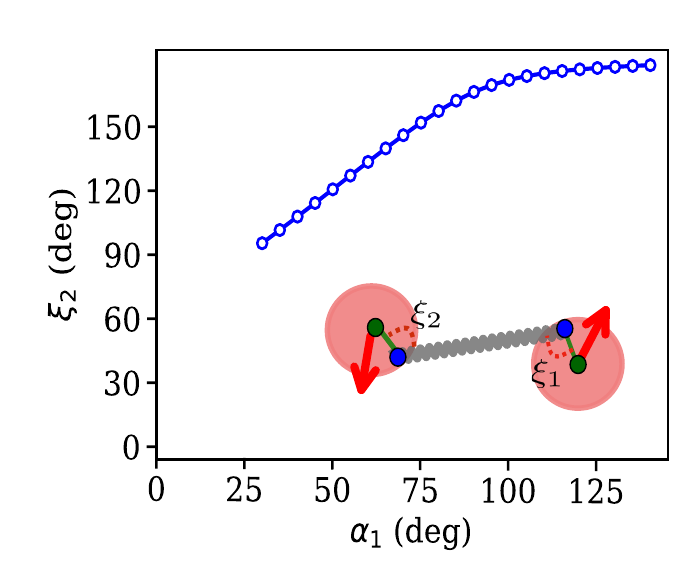}
\end{minipage}
\hfill
\begin{minipage}{0.49\columnwidth}
    \centering
    {\footnotesize{(b)}}
    \includegraphics[width=\linewidth]{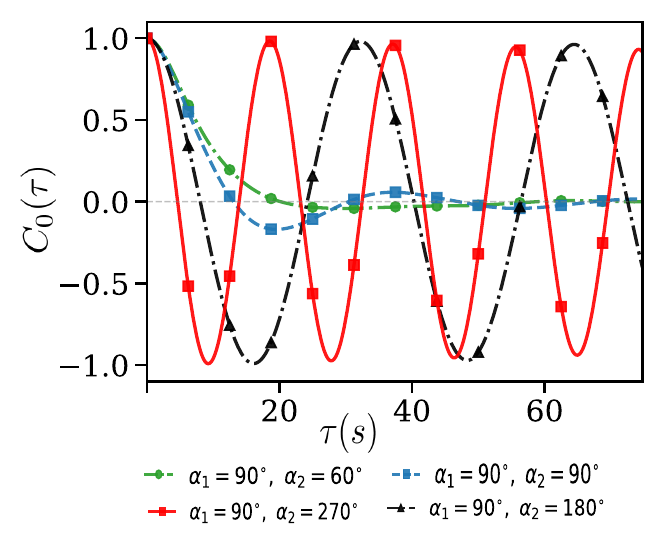}
\end{minipage}
\caption{{(a)} Steady-state angle $\xi_i$ as a function of $\alpha_1$ for $\alpha_1 + \alpha_2 = 360^\circ$. The inset shows a schematic of the system. Here, $\xi_i$ is the angle between the pivot-to-center direction $\mathbf{s}_i$ and the spring vector $\mathbf{r}_s$. (b) Orientational autocorrelation function $C_0(\tau)$ for $\alpha_1 = 90^\circ$ with varying $\alpha_2$. Persistent oscillations indicate spinning motion; faster oscillations correspond to higher $\omega_s$.}
\label{autoxi}
\end{figure}

Fig.~\ref{fig:combined_plots}(b) shows $\langle \omega_s \rangle$ vs $\alpha_1$ along three anti-diagonal lines $\alpha_1 + \alpha_2 = 360^\circ$, $270^\circ$, and $180^\circ$. Along all three lines, $\langle \omega_s \rangle$ exhibits pronounced extrema at intermediate angles, with the sign change indicating a reversal of the spinning direction upon interchanging $\alpha_1$ and $\alpha_2$. The maximum value of $\langle \omega_s \rangle$ is largest along the line $\alpha_1 + \alpha_2 = 360^\circ$, indicating that configurations on this diagonal favor the strongest spinning motion.
For $\alpha_1+\alpha_2=360^\circ$, the steady-state angle $\xi_i$ between $\mathbf{s}_i$ and $\mathbf{r}_s$ decreases as $\alpha_1$ deviates from $180^\circ$. This reduction in $\xi_i$ leads to faster spinning of the system. Figure~\ref{autoxi}(a) shows that $\xi_i$ decreases with decreasing $\alpha_1$; for $\alpha_1<30^\circ$, no stable fixed point is observed. Moreover, the stability of the configuration decreases in the same pattern, such that for $|\alpha_1-180^\circ| \gtrsim 130^\circ$ (i.e., $\alpha_1 \lesssim 50^\circ$ 
or $\alpha_1 \gtrsim 310^\circ$), even small noise drives the system away from the stable fixed point, which in turn reduces the angular speed. The effect of noise is discussed in detail in Sec.~\ref{noise}. 

We now calculate the orientational autocorrelation function,
\begin{equation}
C_0(\tau) = \left\langle \cos\!\bigl[\beta(t+\tau) - \beta(t)\bigr] \right\rangle,
\end{equation}
where $\beta(t)$ is the orientation angle of the vector $\mathbf{r}_s$.  Fig.~\ref{autoxi} (b) shows $C_0(\tau)$ for varying $\alpha_2$ at $\alpha_1 = 90^\circ$. Persistent oscillations in $C_0(\tau)$ directly reflect the periodic rotation of the robotic system about its centroid, i.e., spinning motion. Both $\alpha_2 = 270^\circ$ and $180^\circ$ exhibit clear oscillations in $C_0(\tau)$, confirming spinning motion in both cases. However, the oscillation frequency for $\alpha_2 = 270^\circ$ is notably higher than that for $\alpha_2 = 180^\circ$, indicating a larger $\omega_s$ for the former configuration. In contrast, for $\alpha_2 = 60^\circ$ and $\alpha_1 = 90^\circ$, $C_0(\tau)$ decays rapidly to zero with negligible oscillations, consistent with the negligibly small value of $\omega_s $ for this case.

\begin{figure}[t]
\centering
\begin{minipage}{0.49\columnwidth}
    \centering
     {\footnotesize{(a)}}
    \includegraphics[width=\linewidth]{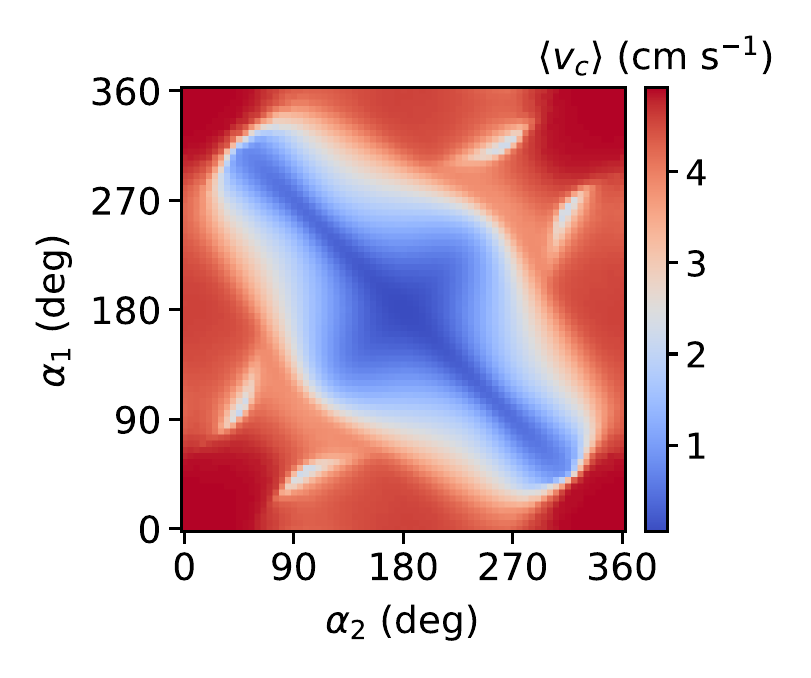}
\end{minipage}
\hfill
\begin{minipage}{0.49\columnwidth}
    \centering
    {\footnotesize{(b)}}
    \includegraphics[width=\linewidth]{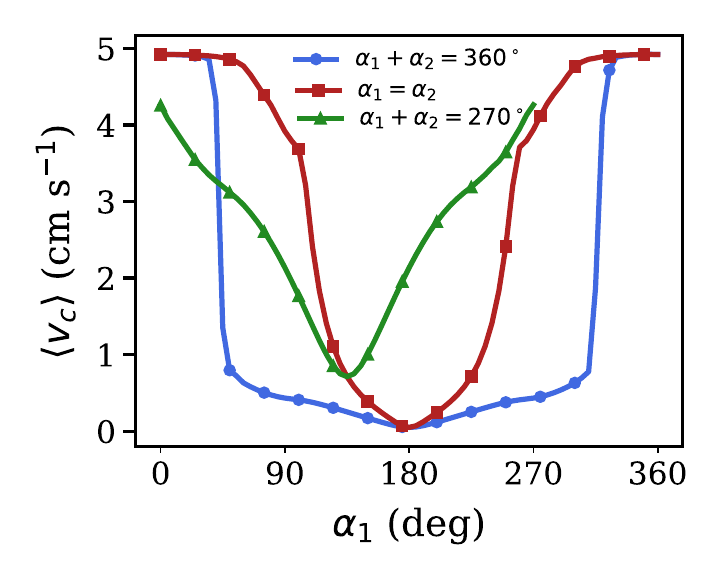}
\end{minipage}
\caption{(a) Heatmap of the average centroid speed $\langle v_c \rangle$ across the $(\alpha_1, \alpha_2)$ parameter space. (b) $\langle v_c \rangle$ along three diagonal cuts: $\alpha_1 + \alpha_2 = 360^\circ$, $\alpha_1 = \alpha_2$, and $\alpha_1 + \alpha_2 = 270^\circ$. All three show suppressed translational motion, with a sharp minimum near $\alpha_1 = 180^\circ$.
}
\label{fig:vel_heatmap}
\end{figure}

The heatmap of the  average centroid speed $\left\langle v_c\right\rangle$ is shown in Fig.~\ref{fig:vel_heatmap}(a). In configurations where the propulsion forces are opposite or nearly opposite—such as near $\alpha_1 = \alpha_2 = 180^\circ$—translational motion is strongly suppressed due to direct geometric cancellation of the propulsive forces. In asymmetric configurations, translation is further suppressed through a distinct mechanism, in which torques redirect the propulsive input into rotation rather than forward motion. In contrast, configurations away from both the force-cancellation and torque-dominated regimes exhibit large centroid speeds, indicating efficient collective translation where the geometric arrangement allows propulsive input to be converted primarily into forward motion.

Fig.~\ref{fig:vel_heatmap}(b) shows broad minima in $\left\langle v_c\right\rangle$ along all three cuts ($\alpha_1 + \alpha_2 = 360^\circ$, $\alpha_1 = \alpha_2$, and $\alpha_1 + \alpha_2 = 270^\circ$), with the sharpest suppression near $\alpha_1 = 180^\circ$, indicating that at this configuration the system exhibits negligible translational motion. The most strongly suppressed translational motion occurs along $\alpha_1+\alpha_2=360^\circ$, consistent with the spinning regime where propulsion is converted into rotation, and for configurations in which the propulsion directions of the two robots are antialigned or nearly so, the active forces cancel, resulting in reduced or vanishing centroid speed.

\subsubsection{Circular motion}
\begin{figure}[htbp]
\centering
\begin{minipage}{0.49\columnwidth}
    \centering
    {\footnotesize{(a)}}\\
    \includegraphics[width=\linewidth]{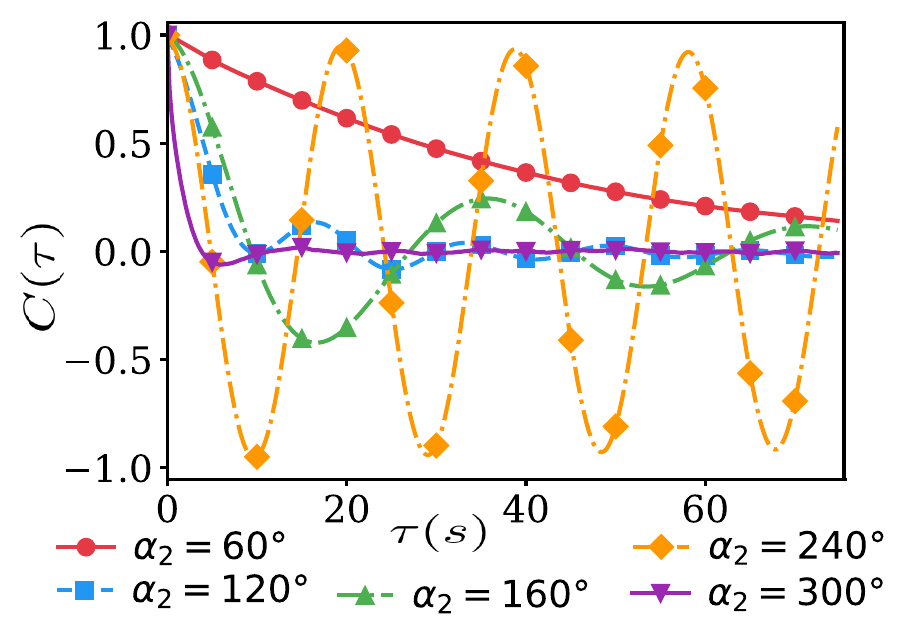}
\end{minipage}
\hfill
\begin{minipage}{0.49\columnwidth}
    \centering
    {\footnotesize{(b)}}\\
    \includegraphics[width=\linewidth]{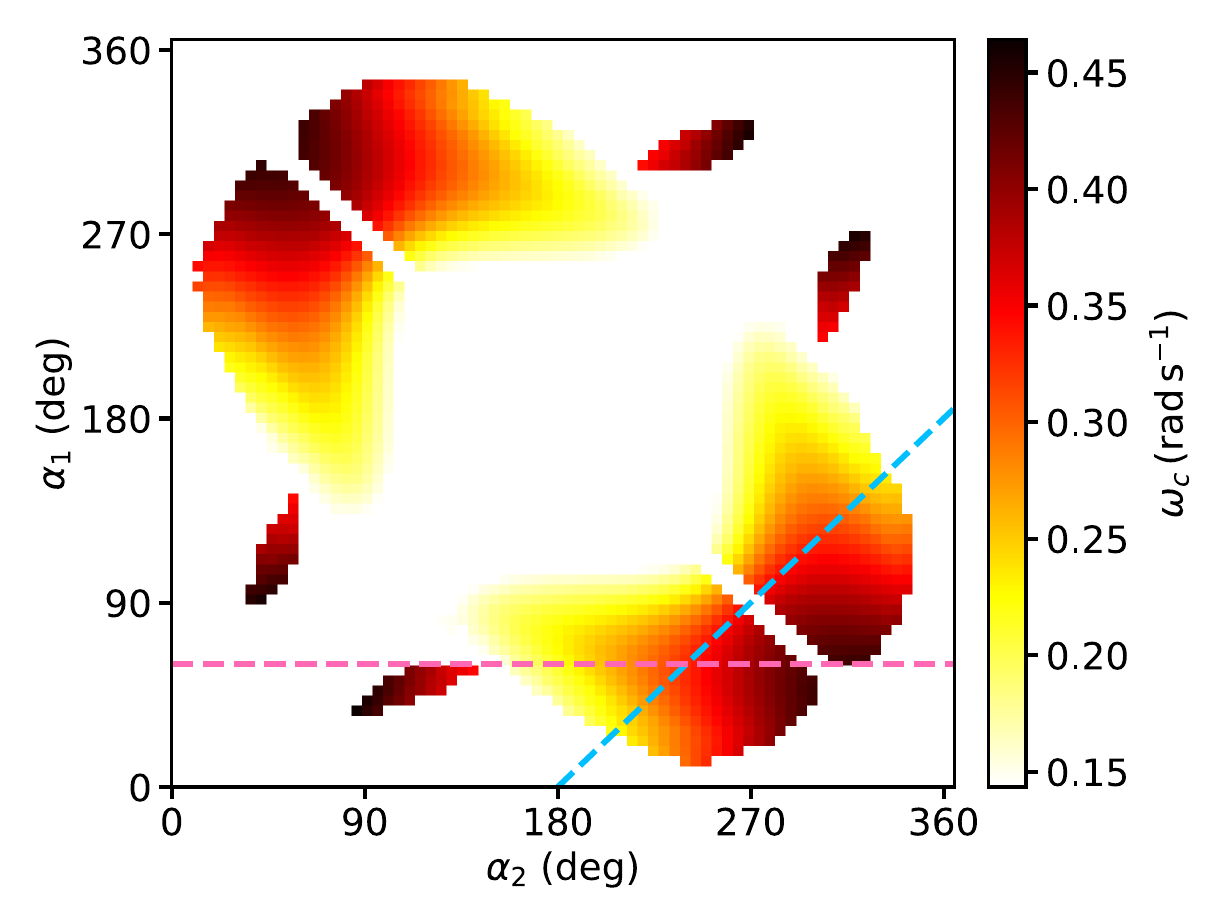}
\end{minipage}
\caption{(a) Autocorrelation function $C(\tau)$ for $\alpha_1 = 60^\circ$ with varying $\alpha_2$, showing pronounced oscillations for configurations exhibiting circular motion. 
(b) Heatmap of the orbital angular velocity $\omega_c$ in the $(\alpha_1,\alpha_2)$ parameter space. Colored regions denote configurations with oscillatory $C(\tau)$, corresponding to circular trajectories, while white regions indicate non-circular motion.}
\label{fig:circular_60}
\end{figure}

We now discuss the circular motion executed by the centroid of the system. To quantify this behavior, we analyze the orientation autocorrelation of the centroid velocity vector $\mathbf{v}_{\mathrm{c}}(t)$, defined as
\[
C(\tau) = \left\langle \cos[\zeta(t+\tau) - \zeta(t)] \right\rangle,
\]
where $\zeta(t)$ is the orientation angle of $\mathbf{v}_{\mathrm{c}}(t)$.
For configurations exhibiting circular motion, the velocity vector $\mathbf{v}_{\mathrm{c}}(t)$ rotates periodically in time, leading to an oscillatory behavior of $C(\tau)$. In contrast, for non-circular motion, $\zeta(t)$ fluctuates irregularly, and $C(\tau)$ decays monotonically without oscillations.
As an example, Fig.~\ref{fig:circular_60}(a) shows $C(\tau)$ for $\alpha_1 = 60^\circ$ with varying $\alpha_2$. The configurations with $\alpha_2 = 240^\circ$ and $160^\circ$ exhibit oscillatory behavior in $C(\tau)$, indicating the presence of circular motion. In contrast, for $\alpha_2 = 300^\circ$ and $60^\circ$, $C(\tau)$ decays to zero without noticeable oscillations, indicating the absence of circular motion.
To extract the orbital angular velocity $\omega_c$ for a given $(\alpha_1,\alpha_2)$, we fit $C(\tau)$ to a damped cosine function of the form.
\[
A e^{-\tau/\tau_d}\cos(\omega_c \tau).
\]
The resulting heat map of $\omega_c$ in the $\alpha_1$--$\alpha_2$ plane is shown in Fig.~\ref{fig:circular_60}(b). The white regions indicate configurations for which negligible oscillatory autocorrelation is detected, corresponding to non-circular motion. Similar to the spinning motion, circular motion is therefore confined to four distinct lobes in the $(\alpha_1,\alpha_2)$ parameter space. When $\alpha_1 + \alpha_2 = 360^\circ$, the net force on the system vanishes. In this case, the centroid undergoes purely diffusive motion. Therefore, no circular motion is observed along this line.
The chirality of the circular orbit is determined by the sign of $\alpha_1-\alpha_2$: positive values correspond to clockwise and negative values to counterclockwise rotation (see movies~\href{https://drive.google.com/drive/folders/1ZZfnyFyMga9KfV9fWmAVixl5AZbw6R6D?usp=sharing}{S8} and~\href{https://drive.google.com/drive/folders/1ZZfnyFyMga9KfV9fWmAVixl5AZbw6R6D?usp=sharing}{S9} for representative clockwise and counterclockwise circular trajectories at $(\alpha_1,\alpha_2)=(240^\circ,60^\circ)$ and $(60^\circ,240^\circ)$, respectively).

\begin{figure}
    \centering
    \includegraphics[width=0.80\linewidth]{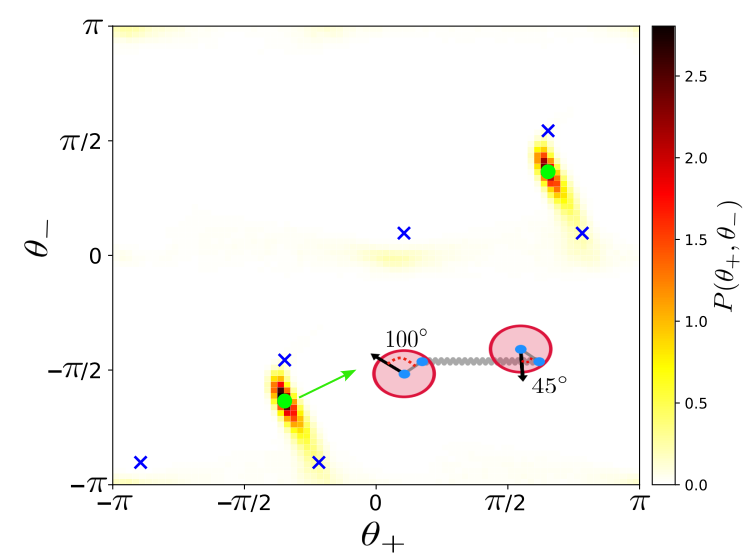}
    \caption{Steady-state probability distribution $P(\theta_+, \theta_-)$ 
    for $(\alpha_1, \alpha_2) = (45^\circ, 100^\circ)$ at 
    $D_r = 0.02~\mathrm{rad}^2\mathrm{s}^{-1}$. The distribution remains localized near the stable fixed point (green dot), with the corresponding robot configuration shown in the inset. This configuration generates a net torque that drives clockwise circular motion.}
    \label{fig:prob_45_100}
\end{figure}

Furthermore, we observe four small regions where the chirality of motion is opposite to that dictated by the sign of $\alpha_1-\alpha_2$, appearing near the boundaries of the dominant circular and spinning lobes in the $(\alpha_1,\alpha_2)$ parameter space [Fig.~\ref{fig:combined_plots}(a) and Fig.~\ref{fig:circular_60}(b)] (see movie~\href{https://drive.google.com/drive/folders/1ZZfnyFyMga9KfV9fWmAVixl5AZbw6R6D?usp=sharing}{S10}). To understand this reversal, we examine the steady-state probability distribution $P(\theta_+,\theta_-)$ for $(\alpha_1,\alpha_2)=(45^\circ,100^\circ)$ at $D_r=0.02~\mathrm{rad}^2\mathrm{s}^{-1}$ [Fig.~\ref{fig:prob_45_100}]. The distribution is localized near a stable fixed point, the configuration of which is illustrated in the inset. At this fixed point, the relative orientations of the two robots are such that the net torque acts in the direction opposite to that of the surrounding parameter space, reversing the chirality of the resulting motion despite the same sign of $\alpha_1-\alpha_2$.

\begin{figure}[htbp]
\centering

\includegraphics[width=0.48\columnwidth]{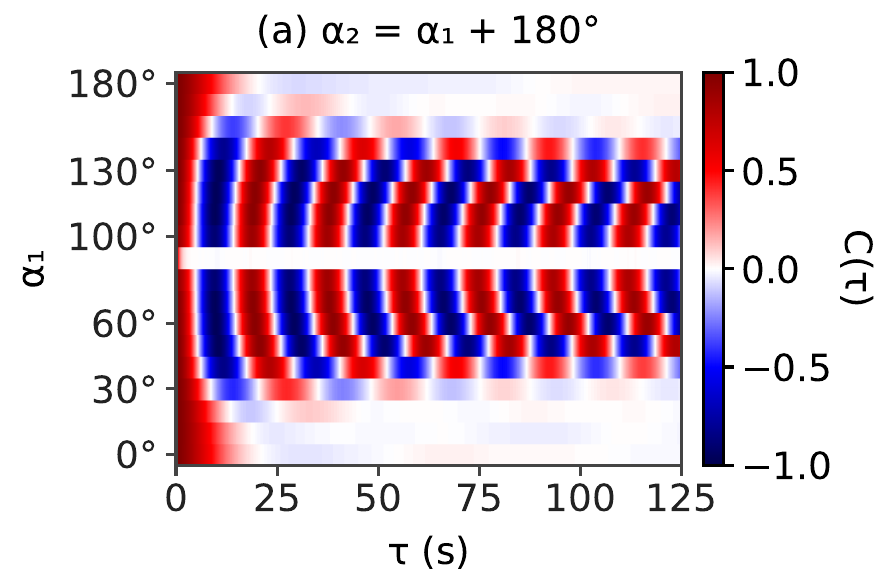}
\hfill
\includegraphics[width=0.48\columnwidth]{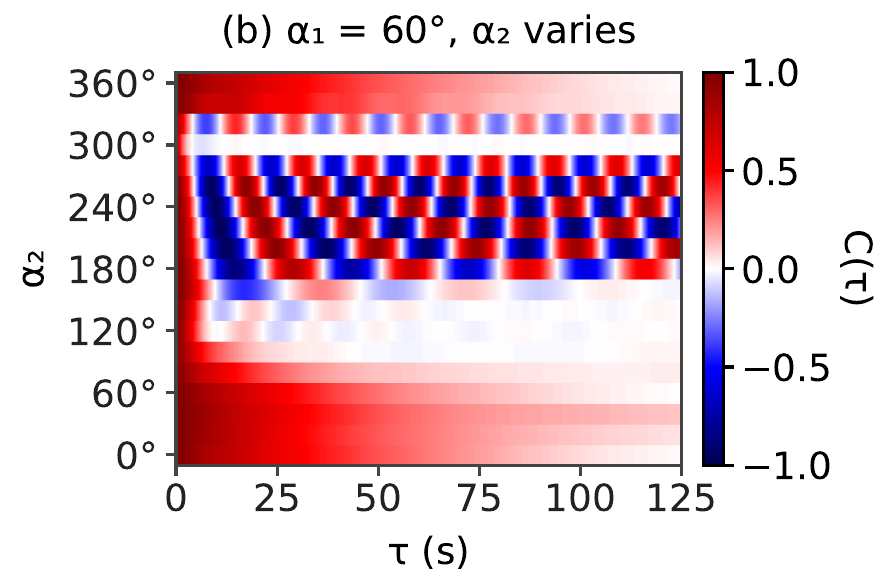}
\includegraphics[width=0.48\columnwidth]{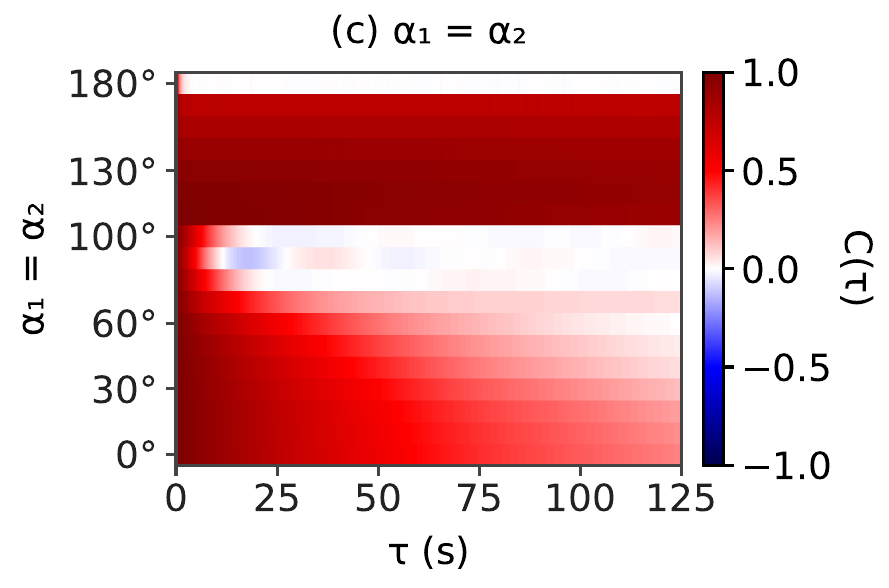}
\hfill
\includegraphics[width=0.48\columnwidth]{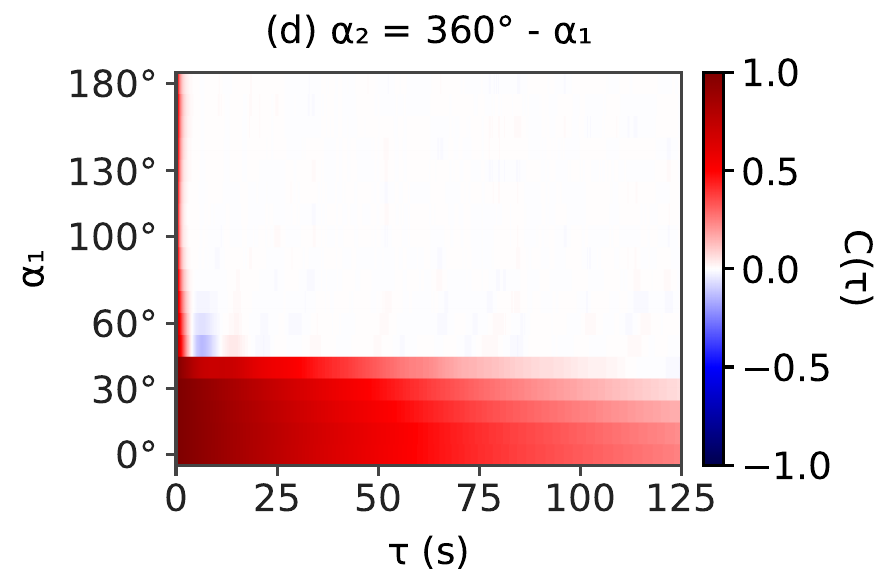}

\caption{Autocorrelation function $C(\tau)$ in the $(\alpha_1,\tau)$ plane for four configurational regimes: (a) Regime $\alpha_2=\alpha_1+180^\circ$, showing persistent oscillations for nearly all $\alpha_1$ values except near $\alpha_1=0^\circ$ and $\alpha_1=180^\circ$; (b) fixed $\alpha_1=60^\circ$ with varying $\alpha_2$, identifying the boundary between circular and non-circular motion; (c) symmetric regime $\alpha_1=\alpha_2$, confirming purely translational motion with no circular component; and (d) $\alpha_2=360^\circ-\alpha_1$, consistent with the spinning regime.}
\label{fig:acf_heatmap}
\end{figure}

Fig.~\ref{fig:acf_heatmap} shows $C(\tau)$ as a heatmap in the $(\alpha_1,\tau)$ plane for four configurational regimes. For the line $\alpha_2=\alpha_1+180^\circ$ [skyblue line Fig.~\ref{fig:circular_60}(b)], persistent oscillations are observed over almost the entire $\tau$ range for nearly all values of $\alpha_1$, except in the regions near $\alpha_1=0^\circ$ and $\alpha_1=180^\circ$ [Fig.~\ref{fig:acf_heatmap}(a)]. This confirms that nearly every configuration along this line exhibits sustained circular motion. For fixed $\alpha_1=60^\circ$ with $\alpha_2$ varying [pink line Fig.~\ref{fig:circular_60}(b)], oscillations appear only for $300^\circ\gtrsim\alpha_2\gtrsim150^\circ$ [Fig.~\ref{fig:acf_heatmap}(b)], directly identifying the boundary between circular and non-circular motion in the parameter space. The symmetric line $\alpha_1=\alpha_2$ shows no oscillations [Fig.~\ref{fig:acf_heatmap}(c)], confirming the purely translational run-and-tumble character with no circular component. For the line $\alpha_2=360^\circ-\alpha_1$, $C(\tau)$ decays rapidly without oscillations [Fig.~\ref{fig:acf_heatmap}(d)], consistent with the spinning regime in which the centroid undergoes nearly diffusive motion.

We now focus on the condition $\alpha_2=\alpha_1+180^\circ$. At small $\alpha_1$, the two robots are nearly co-aligned, producing a large net propulsive force with negligible torque, resulting in weakly curved trajectories with large radii (see movies~\href{https://drive.google.com/drive/folders/1ZZfnyFyMga9KfV9fWmAVixl5AZbw6R6D?usp=sharing}{S11} and~\href{https://drive.google.com/drive/folders/1ZZfnyFyMga9KfV9fWmAVixl5AZbw6R6D?usp=sharing}{S12}). As $\alpha_1$ increases, the robots become progressively more misaligned, reducing the net propulsive force while simultaneously increasing the net torque, which continuously tightens the circular orbits. At $\alpha_1=90^\circ$, the propulsion directions are anti-aligned ($\alpha_1=- \alpha_2$) along $\mathbf{r}_{s\perp}$, the direction perpendicular to the spring, such that the net translational force vanishes while a net torque is generated about the spring axis.
The quantities $\omega_c$, $\langle v_c \rangle$, and $R$ along this line are all symmetric about $\alpha_1 = 90^\circ$, arising from the invariance of the relative geometric configuration under the transformation $\alpha_1 \rightarrow 180^\circ - \alpha_1$, which leaves the magnitude of the net active force and torque unchanged.

\begin{figure}[htbp]
\centering
\includegraphics[width=\linewidth,height=3cm]{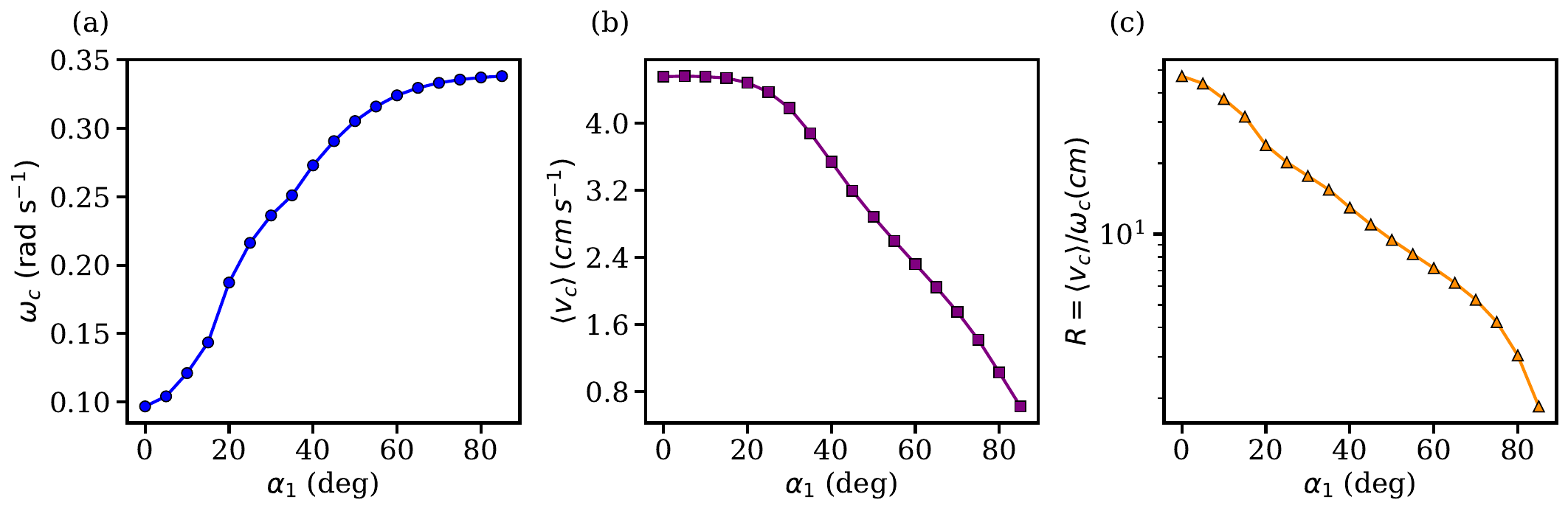}
\caption{(a) Orbital angular velocity $\omega_c$, (b) average centroid speed $\langle v_c \rangle$, and (c) radius of curvature $R = {\langle v_c \rangle}/{\omega_c }$ for $\alpha_2 = \alpha_1 + 180^\circ$.}
\label{wvr}
\end{figure}

As evident from Fig.~\ref{wvr}(a), $\omega_c$ increases monotonically with $\alpha_1$ along line $\alpha_2=\alpha_1+180^\circ$. Simultaneously, $\langle v_c \rangle$ decreases continuously with increasing $\alpha_1$ [Fig.~\ref{wvr}(b)], as the propulsion directions of the two robots become gradually more misaligned, leading to stronger cancellation of the net propulsive force. As a result, the radius of curvature $R=\langle v_c \rangle/\omega_c$ decreases with $\alpha_1$ [Fig.~\ref{wvr}(c)]. As $\alpha_1 \to 90^\circ$, $\langle v_c \rangle$ vanishes while $\omega_c$ remains finite, driving $R \to 0$ and marking the crossover from circular motion to pure spinning.

\vspace{-1em}

\subsubsection{Effect of the rotational noise} \label{noise}
\begin{figure}[htbp]
\centering
\includegraphics[width = 0.80\columnwidth]{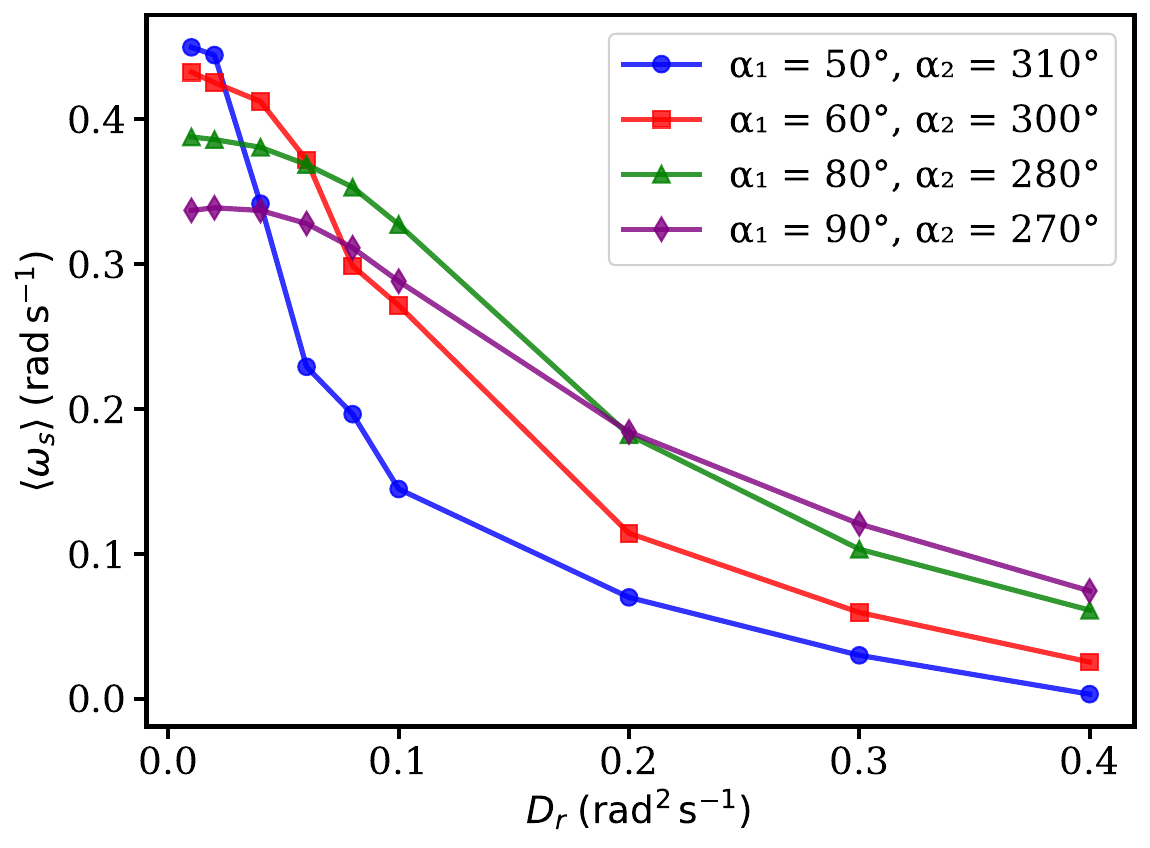}
\caption{Dependence of the average spin angular velocity $\langle \omega_s\rangle$ on the rotational diffusion coefficient $D_r$.}
\label{fig:ws_Dr}
\end{figure}

Here, we study the effect of rotational noise of the robots, characterized by $D_r$, on the spinning and circular motion of the system. Fig.~\ref{fig:ws_Dr} shows the average spin angular velocity $\langle \omega_s \rangle$ as a function of $D_r$ for four different configurations, chosen from the parameter range $50^\circ \le \alpha_1 \le 90^\circ$ along $\alpha_1+\alpha_2=360^\circ$, where the spinning motion is most prominent. As $D_r$ increases, $\langle \omega_s \rangle$ decreases systematically and approaches zero. This occurs because stronger rotational noise randomizes the orientations of the robots, thereby weakening the effect of the geometric asymmetry responsible for the spinning motion. In addition, large noise fluctuations drive the system away from the stable fixed point in the $\theta_--\theta_+$ plane, causing it to transition toward the line associated with the run state. 

\begin{figure}[htbp]
\centering
\includegraphics[width= 1.0\columnwidth]{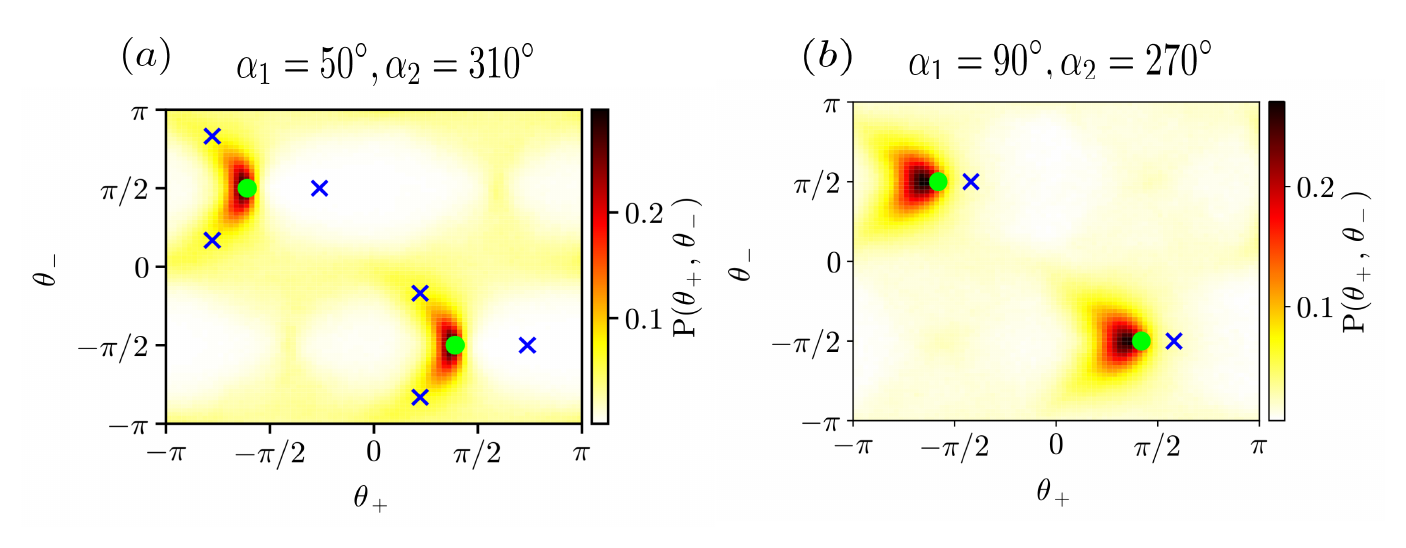}
\caption{Steady-state probability distribution $P(\theta_+, \theta_-)$ for $(\alpha_1, \alpha_2) = (50^\circ, 310^\circ)$ and $(90^\circ, 270^\circ)$ at $D_r = 0.30~\mathrm{rad}^2,\mathrm{s}^{-1}$. For $\alpha_1 = 50^\circ$, the distribution shows a prominent spread toward the run-state region (yellow), indicating frequent noise-driven escape from the spinning fixed point. For $\alpha_1 = 90^\circ$, the distribution remains localized near the stable fixed point with less weight in the run-state region, reflecting greater robustness against rotational noise.}
\label{fig:prob_dist}
\end{figure}

\begin{figure*}[!t]
\centering
\includegraphics[width=0.80\linewidth, height=8.0cm]{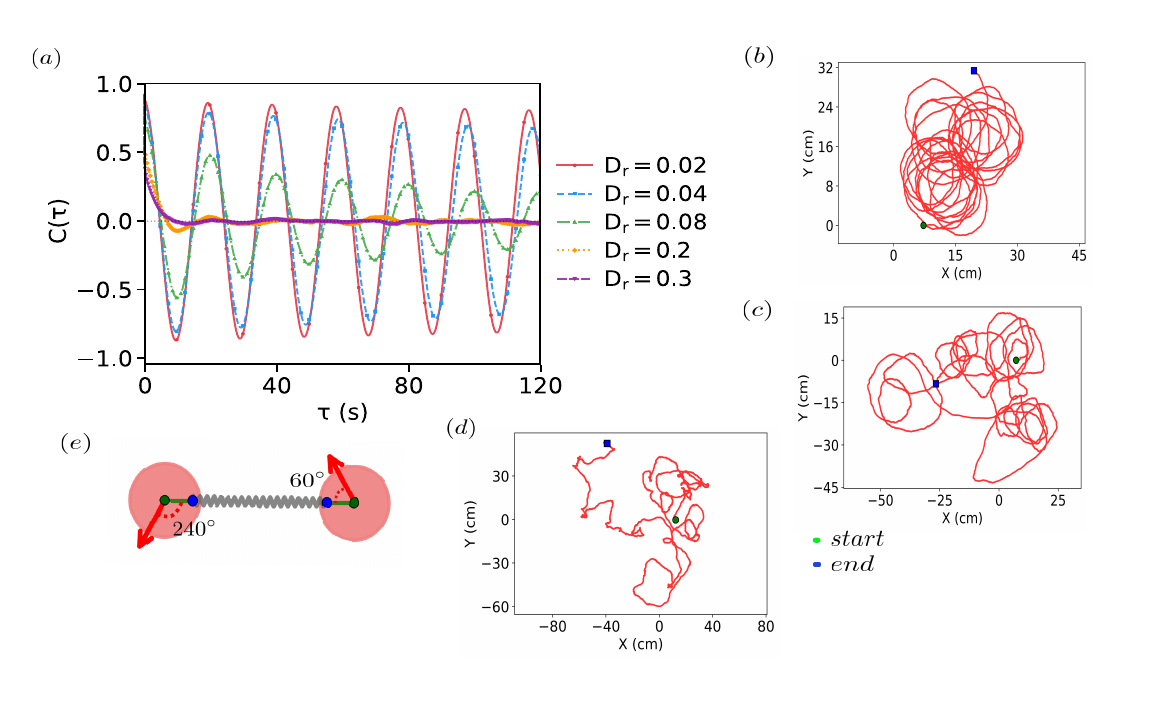}
\caption{(a) Orientational autocorrelation function $C(\tau)$ for the configuration $(\alpha_1,\alpha_2)=(60^\circ,240^\circ)$ at different values of rotational noise $D_r$. 
Trajectories of the system are shown for (b) $D_r=0.02$, (c) $D_r=0.08$, and (d) $D_r=0.30~\mathrm{rad}^2\,\mathrm{s}^{-1}$. 
(e) Schematic configuration of the two-robot system corresponding to $(\alpha_1,\alpha_2)=(60^\circ,240^\circ)$.
}
\label{fig:acf_60_240_allDr}
\end{figure*}

The rate of this suppression depends strongly on the configuration. For small $\alpha_1$, $\langle \omega_s \rangle$ decreases more rapidly with increasing $D_r$. Interestingly, in the large-$D_r$ regime, $\langle \omega_s \rangle$ increases with $\alpha_1$, in contrast to the behavior observed at small $D_r$. This occurs because configurations with larger $\alpha_1$ remain more robust against rotational noise, whereas configurations with smaller $\alpha_1$ are destabilized more easily by the noise of the robots. We examine the steady-state probability distribution $P(\theta_+, \theta_-)$ for two configurations, $(\alpha_1, \alpha_2) = (50^\circ, 310^\circ)$ and  $(\alpha_1, \alpha_2) = (90^\circ, 270^\circ)$ at $D_r=0.30~\mathrm{rad}^2\,\mathrm{s}^{-1}$. For $\alpha_1 = 50^\circ$, the probability is broadly spread across the $\theta_-$ axis, indicating that noise drives the system away from the spinning fixed point toward the run state  yellow region in Fig.~\ref{fig:prob_dist}(a). Conversely, for $\alpha_1 = 90^\circ$, the distribution remains more localized near the stable fixed point, reflecting that this configuration is more robust against rotational noise shown in Fig.~\ref{fig:prob_dist}(b). 

\begin{figure}[t]
\centering
\includegraphics[width=0.80\columnwidth]{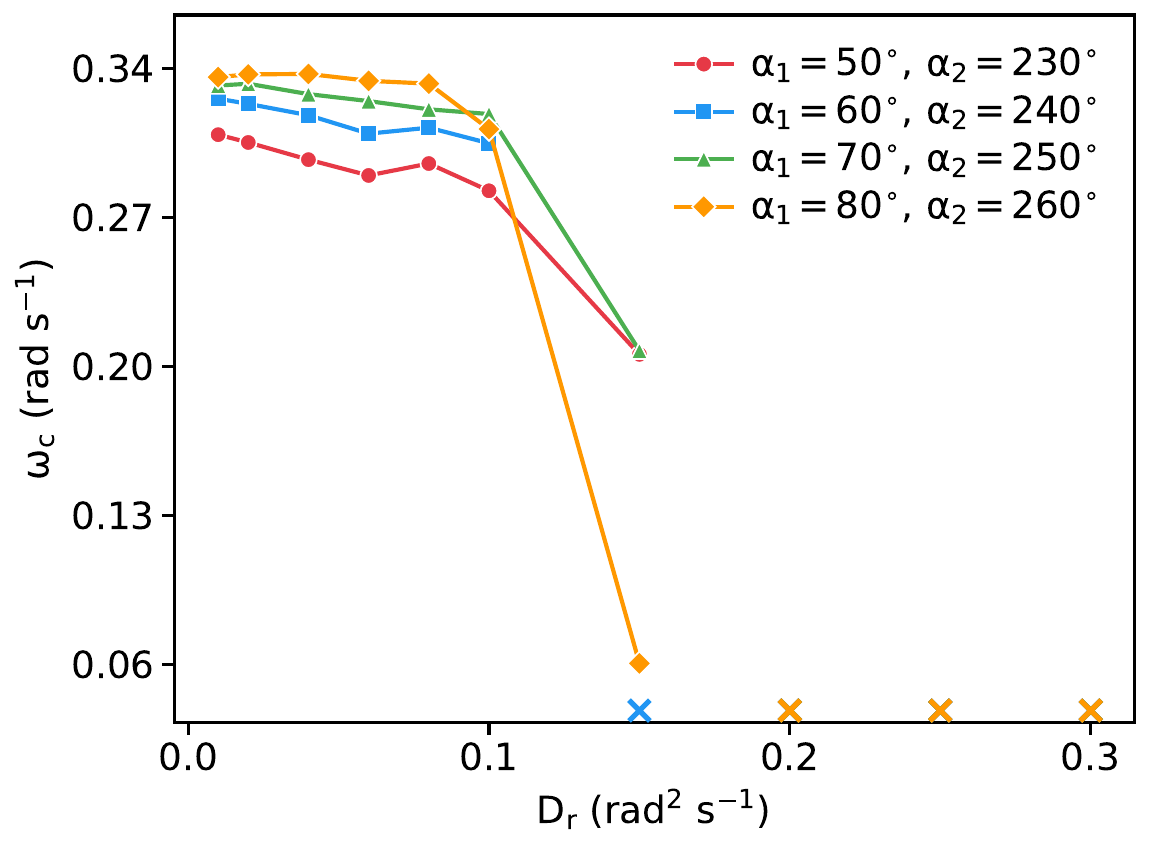}
\caption{Orbital angular velocity $\omega_c$ as a function of rotational noise $D_r$ for $\alpha_2 = \alpha_1 + 180^\circ$. Crosses ($\times$) indicate the absence of circular motion.}
\label{fig:omegac_vs_Dr}
\end{figure}

The autocorrelation function $C(\tau)$ for the $(60^\circ, 240^\circ)$ configuration is shown in Fig.~\ref{fig:acf_60_240_allDr}(a) for increasing $D_r$ values. At $D_r = 0.02~\mathrm{rad}^2\,\mathrm{s}^{-1}$, $C(\tau)$ oscillates with sustained amplitude, reflecting the coherent repeated loops visible in  Fig.~\ref{fig:acf_60_240_allDr}(b). At $D_r = 0.08~\mathrm{rad}^2\,\mathrm{s}^{-1}$, the oscillations persist but with reduced amplitude, mirroring the fewer and more irregular loops seen in Fig.~\ref{fig:acf_60_240_allDr}(c). At $D_r = 0.2$ and $0.3~\mathrm{rad}^2\,\mathrm{s}^{-1}$, the oscillations vanish and $C(\tau)$ decays to zero, in agreement with the near-featureless trajectory of Fig.~\ref{fig:acf_60_240_allDr}(d). 

Fig.~\ref{fig:omegac_vs_Dr} shows the orbital angular velocity $\omega_c$ as a function of $D_r$ for four configurations along $\alpha_2 = \alpha_1 + 180^\circ$. Below a threshold value, $D_r \approx 0.15~\mathrm{rad}^2\,\mathrm{s}^{-1}$, all configurations sustain circular motion, with $\omega_c$ remaining nearly constant; configurations with larger $\alpha_1$ exhibit slightly larger values of $\omega_c$. Above this threshold, $\omega_c$ drops sharply to zero, indicating the disappearance of circular motion.

\subsubsection{Effect of the stiffness of the spring}
In the rigid-rod model of Ref.~\cite{Paramanick2025}, the connection between robots was rigid and inextensible, corresponding to the limit $K \rightarrow \infty$. Replacing the rod with a Hookean spring allows us to continuously tune the mechanical compliance from a highly flexible coupling at small $K$ to the rigid-rod limit at large $K$. This compliance-controlled tunability has biological analogs: in \textit{Chlamydomonas reinhardtii}, the two flagella are connected by a distal striated fiber whose elastic stiffness has been identified as essential for flagellar synchronization, with varying stiffness inducing transitions between distinct synchronization modes~\cite{Wan2016, Klindt2017}.

\begin{figure}[htbp]
\centering
\begin{minipage}{0.49\columnwidth}
    \centering
    (a)\\[-0.2em]
    \includegraphics[width=\linewidth]{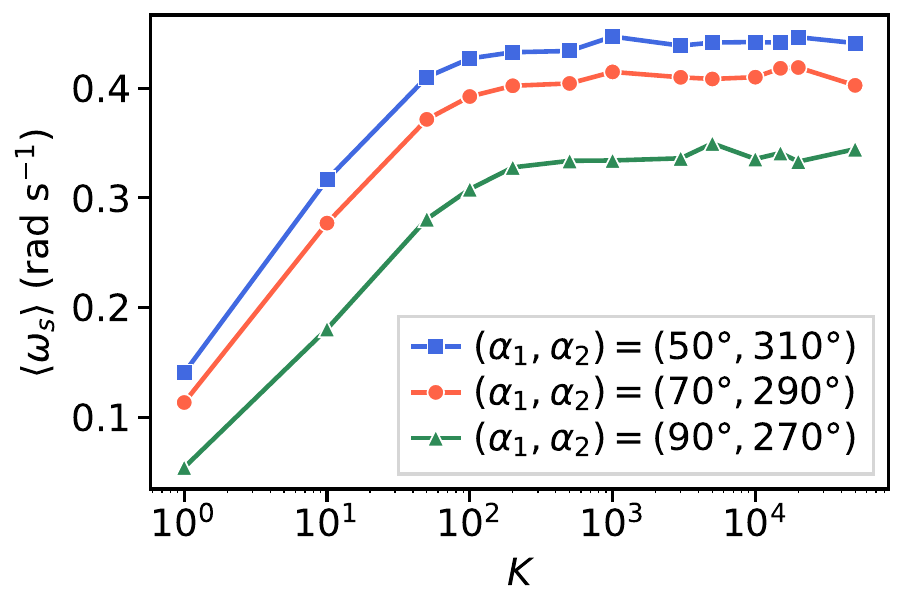}
\end{minipage}
\hfill
\begin{minipage}{0.49\columnwidth}
    \centering
    (b)\\[-0.3em]
    \includegraphics[width=\linewidth]{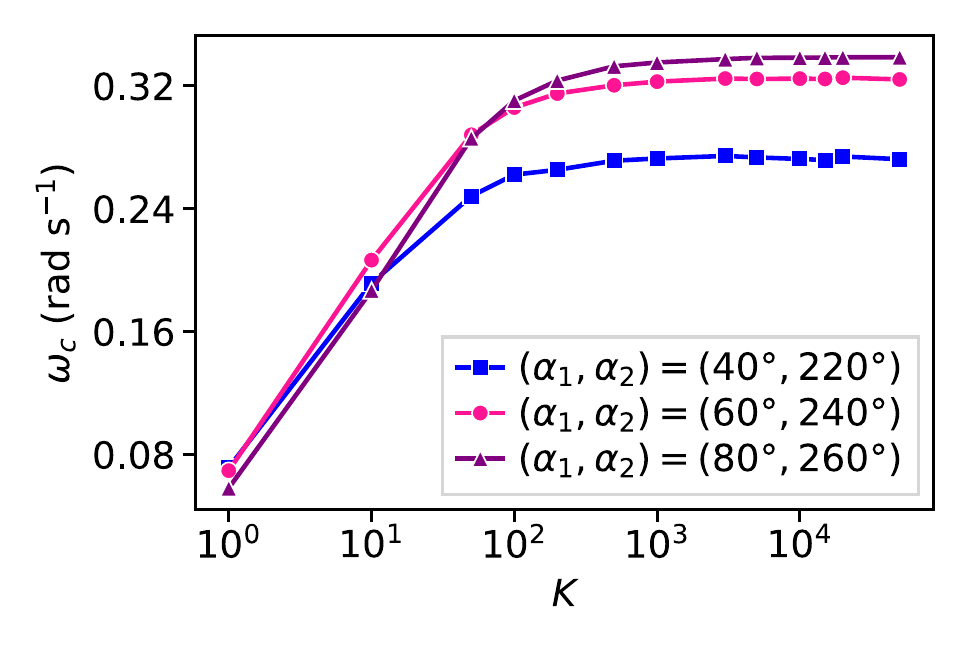}
\end{minipage}
\caption{{(a)} Average spin angular velocity $\langle \omega_s \rangle$ as a function of $K$ for three spinning configurations: $(\alpha_1, \alpha_2) = (50^\circ, 310^\circ)$, $(70^\circ, 290^\circ)$, and $(90^\circ, 270^\circ)$. {(b)} Orbital angular velocity $\omega_c$ as a function of $K$ for three circular-orbit configurations: $(\alpha_1, \alpha_2) = (40^\circ, 220^\circ)$, $(60^\circ, 240^\circ)$, and $(80^\circ, 260^\circ)$. In both panels, velocities increase monotonically with $K$ and saturate at large $K$, approaching the rigid-rod limit.
}
\label{fig:K_dependence}
\end{figure}

We now study how the spring stiffness $K$ affects the spinning and circular motions. Figs.~\ref{fig:K_dependence}(a) and (b) show that both the spin angular velocity $\langle \omega_s \rangle$ and the orbital angular velocity $\omega_c$ increase with $K$ and then saturate, a behavior that is robust across multiple configurations within each motility regime. Specifically, panel~(a) displays $\langle \omega_s \rangle$ for 
three spinning configurations $(\alpha_1, \alpha_2) = (50^\circ, 310^\circ)$, $(70^\circ, 290^\circ)$, and $(90^\circ, 270^\circ)$, while panel~(b) shows $\omega_c$ for three circular configurations $(40^\circ, 220^\circ)$, $(60^\circ, 240^\circ)$, and $(80^\circ, 260^\circ)$. Although the saturated values differ across configurations, reflecting the dependence of the dynamical response on the angles $(\alpha_1$, $\alpha_2$), the qualitative trend is universal. At small $K$, the spring is highly compliant and cannot sustain the constraint force required to transmit sufficient torque between the robots, thereby suppressing the respective dynamical response. As $K$ increases, the spring becomes progressively stiffer, torque transmission between the robots becomes more efficient, and the corresponding dynamical measures grow accordingly. In the limit $K \rightarrow \infty$, the spring effectively becomes inextensible and the system approaches the rigid-rod constraint, leading to saturation of both quantities at their maximum values. 

\subsubsection{Mean square displacement}
\begin{figure}[t]
\centering

\begin{minipage}{0.99\columnwidth}
    \centering
     \includegraphics[width=\linewidth]{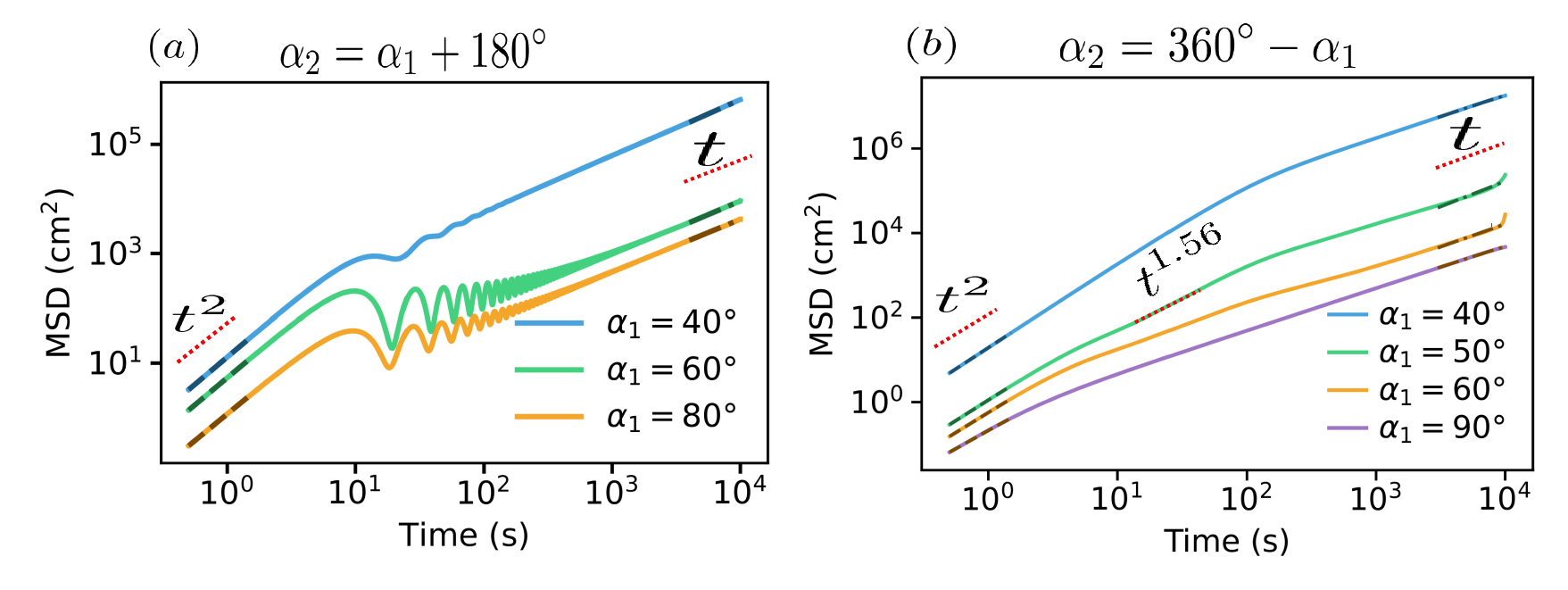}
\end{minipage}

\caption{Mean-square displacement for the (a) circular motion regime ($\alpha_2=\alpha_1+180^\circ$) with $\alpha_1=40^\circ, 60^\circ, 80^\circ$, and (b) spinning regime ($\alpha_2=360^\circ-\alpha_1$) with $\alpha_1=40^\circ, 50^\circ, 60^\circ, 90^\circ$.}
\label{fig:msd}
\end{figure}

Fig.~\ref{fig:msd}   shows the mean-square displacement (MSD) for different configurations across the two motility regimes. In both cases, the MSD grows ballistically, $\mathrm{MSD}\sim t^2$, at short times, reflecting coherent self-propulsion before rotational noise reorients the robots, and crosses over to normal diffusion, $\mathrm{MSD}\sim t$, at long times. The two regimes are, however, clearly distinguished by their intermediate time behavior. For the circular regime, shown in Fig.~\ref{fig:msd}(a) for $\alpha_1 = 40^\circ$, $60^\circ$, and $80^\circ$, the MSD exhibits pronounced oscillations at intermediate times for $\alpha_1 = 60^\circ$ and $\alpha_1 = 80^\circ$, which provide a direct dynamical signature of the underlying circular orbiting. In this regime, the diffusion coefficient decreases strongly with increasing $\alpha_1$, approaching zero as tighter orbits confine the centroid and suppress long-range transport. The configuration $\alpha_1 = 40^\circ$ shows only a weak shoulder in the MSD with no visible oscillations, indicating that at smaller angles $(\alpha_1, \alpha_2)$  the orbital confinement is too loose to produce a measurable periodic signature before diffusion washes it out.

For the spinning regime $\alpha_2 = 360^\circ - \alpha_1$, shown in Fig.~\ref{fig:msd}(b) for $\alpha_1 = 40^\circ$, $50^\circ$, $60^\circ$, and $90^\circ$, the ballistic-to-diffusive crossover is still observed, but the MSD curves are smooth throughout with no oscillations, reflecting the absence of orbital confinement. For the four configurations considered, $\alpha_1 = 40^\circ$ exhibits the longest ballistic persistence, sustained up to $\sim 10^2\mathrm{s}$, before eventually crossing over to normal diffusion. In contrast, for $\alpha_1 = 50^\circ$, the crossover occurs earlier and proceeds through an intermediate superdiffusive regime characterized by $\mathrm{MSD} \sim t^{1.56}$. The remaining configurations, $\alpha_1 = 60^\circ$ and $90^\circ$, transition more rapidly to diffusive behavior, consistent with stronger spinning dynamics and shorter persistence times.

\vspace{-1em}

\section{Discussion and conclusion} ~\label{R&D}
Biological microswimmers such as \textit{Chlamydomonas reinhardtii} propel themselves using two flagella that alternate between synchronized and desynchronized beating states, giving rise to run-and-tumble-like locomotion~\cite{Polin2009}. To explore their environment, these organisms employ diverse motility strategies — from straight runs and tumbles to curved paths, circular swimming, and spinning — driven by the inherent geometric asymmetry between their cis and trans flagella~\cite{Kamiya1987, Cortese2021}. Similar asymmetry-driven chiral motion is observed in \textit{E. coli} swimming in clockwise circles near surfaces~\cite{Lauga2006}.
In this paper, we showed that analogous dynamics arise in a pair of spring-coupled active robots by independently varying the propulsion angles $\alpha_1$ and $\alpha_2$. When $\alpha_1=\alpha_2$, the propulsion geometry is symmetric and the system exhibits run-and-tumble motion, studied in detail in Ref.~\cite{Paramanick2025}. When $\alpha_1\neq\alpha_2$, geometric asymmetry generates a net torque on the system, giving rise to two additional dynamical regimes: spinning motion, in which the torque converts propulsive input into rapid in-place rotation with negligible translation, and circular motion, in which the interplay between translational and rotational dynamics drives the centroid along sustained curved trajectories.

To quantify and distinguish these dynamical regimes, we compute two orientational autocorrelation functions: $C_0(\tau)$, defined from the orientation of the spring vector $\mathbf{r}_s$, and $C(\tau)$, defined from the centroid velocity direction $\mathbf{v}_c$. Persistent oscillations in $C_0(\tau)$ directly reflect the periodic rotation of the system about its centroid, confirming spinning motion, while persistent oscillations in $C(\tau)$ confirm the periodic circulation of the centroid velocity direction, characteristic of circular motion. In the run-and-tumble regime, both $C_0(\tau)$ and $C(\tau)$ decay without oscillations, confirming the absence of persistent rotation.\\
The average centroid speed $\langle v_c \rangle$ also provides a signature of these dynamical regimes. It attains its maximum value for symmetric configurations ($\alpha_1=\alpha_2$), where the propulsive forces of the two robots add constructively, resulting in efficient translational motion characteristic of the run-and-tumble regime. We note, however, that even along the symmetric line $\alpha_1=\alpha_2$, the average speed exhibits a minimum at $\alpha_1=\alpha_2=180^\circ$ and around it. In this configuration, the propulsion directions of the two robots are exactly anti-aligned, leading to a geometric cancellation of the net propulsive force and consequently a strong suppression of translational motion. 
$\langle v_c \rangle$ is strongly suppressed in the spinning regime ($\alpha_1+\alpha_2=360^\circ$), where the net active force vanishes and the propulsive input is entirely converted into in-place rotation with negligible translational motion. In contrast, the circular regime exhibits intermediate values of $\langle v_c \rangle$ that depend on the propulsion angles $(\alpha_1,\alpha_2)$. Along the line $\alpha_2=\alpha_1+180^\circ$, $\langle v_c \rangle$ decreases monotonically with increasing $\alpha_1$. At the same time, the orbital angular velocity $\omega_c$ increases, leading to a reduction in the orbital radius, $R={\langle v_c\rangle}/{\omega_c}$. \\
At low $D_r$, the average spin angular velocity $\langle \omega_s \rangle$ increases with decreasing $\alpha_1$ along the line $\alpha_1+\alpha_2=360^\circ$. As $D_r$ increases, this trend reverses. Configurations with larger $\alpha_1$ maintain higher values of $\langle \omega_s \rangle$ because their orientational dynamics remain more strongly localized around the stable fixed point. By contrast, configurations with smaller $\alpha_1$ are more susceptible to rotational fluctuations, which destabilize the fixed-point localization and reduce $\langle \omega_s \rangle$. 
For the circular regime, increasing $D_r$ progressively disrupts the orbital dynamics. Above a threshold value of approximately $D_r \approx 0.15~\mathrm{rad}^2\mathrm{s}^{-1}$, the orbital angular velocity $\omega_c$ rapidly decreases toward zero, indicating the loss of persistent circular motion. This behavior aligns well with the established picture of chiral active particles, where rotational diffusion perturbs circular trajectories and the decorrelation time $\tau_R = D_r^{-1}$ sets the timescale over which orientational memory is lost~\cite{Caprini2022,Caprini2023}. A similar interplay between intrinsic torque and orientational fluctuations is believed to explain why circular swimming persists in \textit{E.\ coli} near surfaces~\cite{Lauga2006}. \\

The elastic coupling through the spring plays a crucial role in mediating these dynamics. Both $\langle \omega_s \rangle$ and $\omega_c$ increase with $K$ and saturate in the large-$K$ limit, recovering the rigid-rod constraint. In \textit{Chlamydomonas reinhardtii}, the two flagella are connected at their bases by a distal striated fiber that acts as an elastic spring-like coupler. This basal elastic link has been identified as essential for flagellar synchronization~\cite{Wan2016}, and theoretical work has shown that simply tuning the stiffness of this coupling can drive the system from one synchronization mode to another~\cite{Klindt2017}. A similar mechanism occurs in the bacterial flagellar hook, whose flexibility promotes filament bundling in swimming mode, but increased rigidity breaks the bundle apart during tumbling~\cite{Chen2023}.

The mean-square displacement(MSD) further distinguishes the two motility regimes. At short times, both exhibit ballistic growth before rotational noise reorients the robots. In the circular regime, the MSD displays oscillatory behavior at intermediate times, while the long-time diffusivity is strongly suppressed because the centroid remains localized within the orbit. In the spinning regime, the MSD evolves smoothly, with the effective diffusivity decreasing as $\alpha_1$ increases, as more propulsive input is redirected into rotation rather than translation.

These results demonstrate that geometric asymmetry and elastic compliance are sufficient to generate a broad spectrum of emergent motility behaviors --- from straight runs and tumbles to circular trajectories and spinning --- without requiring hydrodynamic interactions, chemical gradients, or active sensing. These findings provide a minimal mechanical design principle applicable to both synthetic active systems with tunable locomotion modes and biological microswimmers, where structural asymmetries can produce diverse motility behaviors through purely mechanical interactions.

\begin{acknowledgments}
N.K. acknowledges financial support from Anusandhan National Research Foundation (ANRF) for Advanced Research Grant No. ANRF/ARG/2025/005689/PS.
\end{acknowledgments}

\bibliography{apssamp}

\end{document}